\documentclass{iopjournal}
\usepackage[T1]{fontenc}
\usepackage{graphicx}
\usepackage{amsmath}
\usepackage{braket}
\usepackage{amsfonts}
\usepackage{bm}
\usepackage{amssymb}
\usepackage{cite}

\newcommand{\dd}{\mathrm{d}}
\newcommand{\ii}{\mathrm{i}}
\newcommand{\E}{\mathbb{E}}
\newcommand{\Reals}{\mathbb{R}}
\newcommand{\ip}[2]{\langle #1 \mid #2 \rangle}
\newcommand{\norm}[1]{\left\lVert #1 \right\rVert}
\newcommand{\pd}[2]{\frac{\partial #1}{\partial #2}}

\begin{document}

\articletype{Paper}

\title{Geometric distinguishability of two-source gravitational-wave signals from the extrinsic geometry of the single-source waveform manifold}

\author{Cezar Valentin Ionescu$^{1,2}$, Paul-Adrian Gogîță$^{1,2}$, Laurențiu Ioan Caramete$^{1, *}$, and Ana Caramete$^1$}

\affil{$^1$Institute of Space Science -- INFLPR Subsidiary, Atomiștilor 409, Măgurele, România}

\affil{$^2$Faculty of Physics, University of Bucharest, Atomiștilor 405, Măgurele, România}

\affil{$^*$Author to whom any correspondence should be addressed.}

\email{lcaramete@spacescience.ro}

\keywords{gravitational waves, parameter estimation, waveform manifolds, Fisher information, extrinsic curvature, source confusion}

\begin{abstract}
Overlapping gravitational-wave signals can mimic a single effective source
when parameter degeneracies are strong. We formulate this source-confusion
problem geometrically by treating the family of single-source waveforms as
an embedded manifold in the noise-weighted detector Hilbert space. The
minimum residual of a single-source fit defines a distance-to-manifold
statistic whose square equals twice the maximized log-likelihood gain of the
two-source hypothesis in the noiseless limit and, in Gaussian noise, is the
deterministic residual power that sets the expected gain of the nested
likelihood-ratio comparison.
When two sources are close in parameter space, optimal single-source template
placement absorbs all variations tangent to the manifold; the leading
irreducible residual lies in the normal bundle and is governed by the second
fundamental form, yielding a quartic scaling of the squared distance with
parameter separation. We verify this scaling using a six-parameter
aligned-spin inspiral model (1.5PN phase, 0.5PN amplitude) coupled to the
LISA orbital response with instrumental and Galactic confusion noise.
Constrained numerical minimization along seven directions across six
reference binaries confirms the predicted quartic scaling, prefactor, and
unequal-amplitude harmonic-mean rule throughout the perturbative regime,
while delineating the boundary of validity along soft directions.
Confusion maps in representative parameter planes reveal strongly anisotropic
boundaries: while generic directions are curvature-limited, exact degeneracies---such
as the 1.5PN spin null line---produce prior-bounded confusion needles. Extrinsic
curvature thereby provides a coordinate-invariant diagnostic that directly maps
which parameter directions permit severe source confusion. The calculations are
implemented in an accompanying open-source Julia package.
\end{abstract}

\section{Introduction}\label{sec:intro}

Matched filtering remains the canonical framework for detecting and characterizing deterministic gravitational-wave (GW) signals in coloured, approximately stationary Gaussian noise. In this setting the data stream is viewed as a vector in a noise-weighted inner-product space, and likelihood comparisons reduce to squared distances induced by the detector power spectral density~\cite{Finn1992,JaranowskiKrolak2005}.
For waveform families parameterized by physical parameters $\theta\in\Theta$, the Fisher information matrix
$g_{ij}(\theta)=\braket{\partial_i h(\theta)| \partial_j h(\theta)}$
gives the local quadratic approximation to the log-likelihood and underpins much of the standard intuition for parameter estimation~\cite{CutlerFlanagan1994}.
The same inner-product geometry also drives the construction of template banks, where a Riemannian ``mismatch metric'' controls the loss in signal-to-noise ratio due to the discreteness of the bank~\cite{Owen1996}.
While powerful, this intrinsic viewpoint does not address a central question in crowded data streams: when can the superposition of multiple astrophysical signals mimic a single effective source from the same model family?

In space-based gravitational-wave detection, overlapping signals are generic rather than exceptional. In the millihertz band targeted by the Laser Interferometer Space Antenna (LISA), the data stream will record tens of thousands of continuous Galactic compact binaries alongside merging massive black-hole binaries and extreme-mass-ratio inspirals~\cite{AmaroSeoaneEtAl2017}. Unresolved or partially resolved compact binaries form a confusion foreground that impedes signal characterization across the band~\cite{TimpanoRubboCornish2006}.
Bayesian and global-fit pipelines have been developed to resolve overlapping signals simultaneously while inferring the confusion noise level~\cite{UmstatterEtAl2005,CrowderCornish2007,Littenberg2011,Littenberg2023,StrubEtAl2024,Katz2025,Deng2025}.
In such crowded regimes, distinguishing competing source counts is frequently limited by geometric near-degeneracies in waveform space rather than by instrumental noise alone.

Standard Fisher-matrix expansions break down near strong correlations or exact degeneracies, where quadratic approximations fail to reflect the likelihood topology~\cite{Vallisneri2008}.
This calls for coordinate-invariant diagnostics grounded in the global geometry of the waveform manifold and sensitive to its extrinsic embedding within the ambient signal space.
Geometrically, the map $h:\Theta\to\mathcal{H}$ from parameters to noise-weighted waveforms defines an embedded manifold of single-source signals,
$\mathcal{M}_1=h(\Theta)\subset\mathcal{H}$,
where $\mathcal{H}$ is the detector Hilbert space induced by the noise inner product~\cite{Finn1992,JaranowskiKrolak2005}.
Multi-source hypotheses define corresponding composite sets, such as
$\mathcal{M}_2=\{h(\theta_1)+h(\theta_2):\theta_1,\theta_2\in\Theta\}\subset\mathcal{H}$.
Resolving multiple sources from a single template is therefore an extrinsic geometric problem in $\mathcal{H}$.

The geometry of nonlinear least squares as a nearest-point problem on a
model manifold, with the curvature of the manifold controlling the
departure from linear behaviour, is classical in
statistics~\cite{BatesWatts1980} and has been developed for sloppy models in
prediction space~\cite{TranstrumMachtaSethna2011}. The statistical
resolution of two closely spaced deterministic signals has been analysed in
array processing and imaging, where the minimum resolvable separation falls
as the inverse fourth root of the signal-to-noise ratio in
power~\cite{Smith2005,ShahramMilanfar2004}, the scaling that the quartic law
below implies in raw parameter coordinates; the biases that an unmodelled
overlapping signal induces in the recovered parameters of another have been
treated at the linear-signal level for gravitational
waves~\cite{AntonelliBurkeGair2021}, and source confusion in the LISA band
was studied directly in~\cite{CrowderCornish2004}. In this paper we determine the
extrinsic geometric tensor governing local two-to-one source
confusion, derive its amplitude scaling, and evaluate it
for an inspiral family under the full orbital LISA response.

We analyze the distance-to-manifold statistic
\begin{equation}
D(d):=\inf_{\theta\in\Theta}\,\|d-h(\theta)\|,
\end{equation}
where $\|\cdot\|$ is the norm induced by the detector noise-weighted inner product.
This statistic has a direct likelihood interpretation: maximizing the single-source likelihood over $\theta$ is equivalent to minimizing $\|d-h(\theta)\|^2$.
In a noiseless baseline, if $d=\omega\in\mathcal{M}_2$ is generated by two sources within the model and the two-source hypothesis fits $\omega$ exactly, the maximized log-likelihood ratio between the two-source and one-source hypotheses reduces to
$\Delta\ln L_{\max}(\omega)=\tfrac12 D^2(\omega)$.
The quantity $D^2(\omega)$ measures the residual power that cannot be absorbed by any single-source template.

Our main analytical result concerns the \emph{local} resolvability of two nearby sources.
The leading-order behavior of $D^2$ is governed not by the intrinsic Fisher metric alone, but by the \emph{second fundamental form} of the embedding $\mathcal{M}_1\hookrightarrow\mathcal{H}$.
When two sources have nearby parameters (under an appropriate amplitude--shape split), optimizing over the effective single-source parameters eliminates all tangent-space components of the mismatch; the irreducible residual lies in the normal space and is governed by the normal projection of second derivatives $\partial_i\partial_j h$.
This yields a quartic scaling law of the schematic form
\begin{equation}
D^2 \;\simeq\; \frac{1}{16}\,\|\mathrm{II}(u,u)\|^2\,\delta^4
\qquad (\delta\to 0),
\end{equation}
where $\delta$ is the separation in parameter space along the unit direction $u$, $\mathrm{II}$ is the second fundamental form of the single-source manifold at the source~\cite{doCarmo1992,Lee2018Riemannian}, and $\mathrm{II}(u,u)$ is its normal curvature vector along $u$.
Geometrically, parameter directions of small extrinsic curvature correspond to weakly curved embeddings in $\mathcal{H}$ that permit severe multi-to-single degeneracies, whereas directions of large normal curvature generate substantial residuals, rendering the sources distinguishable at much smaller separations.

The paper is organised as follows. Section~\ref{sec:noise} reviews the noise-weighted Hilbert space and inner product. In Section~\ref{sec:single} we define the single-source manifold and its second fundamental form~\cite{CutlerFlanagan1994,Owen1996}. Section~\ref{sec:multi} introduces multi-source signal sets and formulates model selection as a distance-to-manifold problem. Section~\ref{sec:local} develops the local perturbative expansion, deriving the quartic resolvability law and its amplitude scaling. In Section~\ref{sec:numerical-validation} we test this scaling against direct numerical minimization for aligned-spin inspirals under the LISA response, presenting one-dimensional parameter sweeps, residual spectra, and confusion maps. Finally, Section~\ref{sec:conclusions} discusses the implications for global-fit analysis pipelines and outlines future directions~\cite{TimpanoRubboCornish2006,CrowderCornish2007,Littenberg2011,Littenberg2023,StrubEtAl2024,Katz2025,Deng2025}.

\section{Statistical model and noise-weighted Hilbert space}\label{sec:noise}

\subsection{Data model and assumptions}

For a detector network or multichannel instrument with $N_c$ read-out channels, the output strain in channel $I\in\{1,\ldots,N_c\}$ over an observation interval $t\in[0,T]$ is modelled as
\begin{equation}
d_I(t)=h_I(t)+n_I(t),
\end{equation}
where $h_I(t)$ is the gravitational-wave strain response and $n_I(t)$ denotes the detector noise. When treating a single channel or generic time series, we suppress the channel index and write $d(t)=h(t)+n(t)$.

We assume the noise $n_I(t)$ is a zero-mean, wide-sense stationary Gaussian random process that is uncorrelated across channels:
\begin{equation}
\E[n_I(t)]=0, \qquad \E[n_I(t)\,n_J(t')]=\delta_{IJ}\,R_I(t-t'),
\end{equation}
where $R_I(\tau)$ is the autocorrelation function of channel $I$.
For discrete time samples $n_k=n(t_k)$ collected in $\bm n\in\Reals^N$, the noise covariance matrix has components $C_{k\ell}=\E[n_k n_\ell]$ and the joint probability density is
\begin{equation}
p(\bm n)=\frac{1}{\sqrt{(2\pi)^N\det C}}\exp\!\left(-\frac12\,\bm n^\mathsf{T}C^{-1}\bm n\right).
\end{equation}
Since $\bm n=\bm d-\bm h$, the likelihood of observing data $\bm d$ given a predicted signal $\bm h$ is
\begin{equation}
p(\bm d\mid \bm h)=\frac{1}{\sqrt{(2\pi)^N\det C}}\exp\!\left(-\frac12(\bm d-\bm h)^\mathsf{T}C^{-1}(\bm d-\bm h)\right).
\end{equation}

For hypothesis testing against noise-only data, one works with the log-likelihood ratio between the signal hypothesis $H_1:\, d=h(\theta)+n$ and the noise-only hypothesis $H_0:\, d=n$:
\begin{equation}
\Lambda(\theta;d):=\ln\frac{p(d\mid h(\theta))}{p(d\mid 0)}
= \braket{d |h(\theta)}-\frac12\braket{h(\theta)|h(\theta)},
\end{equation}
where $\braket{\cdot|\cdot}$ denotes the noise-weighted inner product defined below.
Maximizing $\Lambda$ over an overall amplitude yields the standard matched-filter signal-to-noise ratio (SNR)
\begin{equation}
\rho(\theta)=\frac{\braket{d |h(\theta)}}{\sqrt{\braket{h(\theta)|h(\theta)}}},
\end{equation}
which is optimal in the Neyman--Pearson sense for known waveforms in stationary Gaussian noise~\cite{Finn1992,JaranowskiKrolak2005,CutlerFlanagan1994}.

\subsection{Frequency domain and the noise-weighted inner product}
Because stationarity implies that the time-domain covariance depends only on time differences, $C_{ij}=R_n(t_i-t_j)$, the covariance matrix $C$ for uniformly sampled data is Toeplitz and is asymptotically diagonalized by the discrete Fourier transform (DFT)~\cite{Gray2006Toeplitz}. Letting $F$ denote the DFT unitary matrix,
\begin{equation}
C\simeq F^{-1}\Sigma F,\qquad C^{-1}\simeq F^{-1}\Sigma^{-1}F,
\end{equation}
where $\Sigma$ is diagonal.
For a sampled time series with duration $T=N\Delta t$ and frequencies $f_j=j/T$, the discrete Fourier transform of the noise is
\begin{equation}
\tilde n(f_j)\simeq \Delta t\sum_{k=0}^{N-1} n(t_k)e^{-2\pi \ii f_j t_k}.
\end{equation}

In the continuous-time representation, the autocorrelation $R_n(\tau)=\E[n(t)n(t+\tau)]$ fixes the one-sided power spectral density (PSD) $S_n(f)$ via the Wiener--Khinchin relation:
\begin{equation}
S_n(f)=2\int_{-\infty}^{\infty}R_n(\tau)e^{2\pi \ii f\tau}\,\dd\tau,
\qquad f\ge 0,
\end{equation}
where the factor of $2$ reflects the one-sided convention standard in gravitational-wave astronomy~\cite{JaranowskiKrolak2005,Allen2012Findchirp}.
With the continuous Fourier transform $\tilde n(f)=\int_{-\infty}^{\infty}n(t)e^{-2\pi \ii f t}\,\dd t$,
stationarity yields
\begin{equation}
\E[\tilde n(f)\tilde n^{*}(f')]=\frac12 S_n(|f|)\,\delta(f-f').
\end{equation}
For a finite observation window of duration $T$, the truncated Fourier transform $\tilde n_T(f)=\int_{0}^{T}n(t)e^{-2\pi \ii f t}\,\dd t$ has second moment
\begin{equation}
\E[\tilde n_T(f)\tilde n^{*}_T(f')]
=\int_0^T\!\!\int_0^T R_n(t-t')\,e^{-2\pi \ii (f t-f' t')}\,\dd t\,\dd t'.
\end{equation}
Changing variables to $\tau=t-t'$ and $u=(t+t')/2$, the finite-time covariance takes the form
\begin{equation}
\E[\tilde n_T(f)\tilde n^{*}_T(f')]\simeq
\frac12 S_n(|f|)\,T\,\mathrm{sinc}\!\big[\pi(f-f')T\big]\,e^{-\ii\pi(f-f')T}
\;\xrightarrow{\;T\to\infty\;}\;\frac12 S_n(|f|)\,\delta(f-f'),
\end{equation}
with $\mathrm{sinc}\,x=\sin x/x$. The diagonal variance $\E|\tilde n_T(f)|^2=\tfrac{T}{2}S_n(|f|)$ sets the normalization of the frequency-domain likelihood.

The noise-weighted inner product between real time series $a(t), b(t)$ on $[0,T]$ is defined by
\begin{equation}
\braket{a|b}=4\,\Re\int_0^\infty \frac{\tilde a^{*}(f)\tilde b(f)}{S_n(f)}\,\dd f,
\end{equation}
inducing the Hilbert space norm $\norm{a}^2=\braket{a|a}$.
The detector Hilbert space $\mathcal{H}$ consists of real time series on $[0,T]$ with finite norm, $\norm{a}<\infty$, equipped with $\braket{\cdot|\cdot}$.
For an instrument with $N_c$ uncorrelated channels ($S_{IJ}(f)=\delta_{IJ}S_I(f)$), the inner product generalizes to the direct sum across channels~\cite{JaranowskiKrolak2005,TintoDhurandhar2014}:
\begin{equation}
\braket{a|b}=\sum_{I=1}^{N_c} 4\,\Re\int_0^\infty \frac{\tilde a_I^{*}(f)\tilde b_I(f)}{S_I(f)}\,\dd f.
\end{equation}
For data sampled at cadence $\Delta t$ over duration $T$, the continuous inner product is approximated on the discrete frequency grid $f_k=k/T$ ($\Delta f=1/T$) by
\begin{equation}
\braket{a|b}\;\approx\;4\,\Re\sum_{k=1}^{\lceil N/2\rceil-1}\frac{\tilde a^{*}(f_k)\tilde b(f_k)}{S_n(f_k)}\,\Delta f,
\end{equation}
excluding the zero-frequency bin $k=0$ and, for even $N$, the Nyquist bin.
In this Hilbert-space notation, the Gaussian likelihood and log-likelihood become
\begin{equation}
p(d\mid h)=\frac{1}{\sqrt{(2\pi)^N\det C}}\exp\!\left[-\frac12\norm{d-h}^2\right],
\qquad
\ln p(d\mid h)=-\frac12\norm{d-h}^2+\text{const}.
\end{equation}

\section{Single-source waveform manifold}\label{sec:single}

Let $\Theta$ denote the single-source parameter space, a smooth $m$-dimensional manifold whose coordinates $\theta^i$ encompass intrinsic physical parameters (such as component masses and spins), extrinsic geometry (sky location, inclination, polarization), and reference quantities (distance, coalescence time, phase).
A deterministic gravitational-wave model specifies a mapping
\begin{equation}
h:\Theta\to\mathcal{H},\qquad \theta\mapsto h(\theta),
\end{equation}
assigning to each parameter point $\theta\in\Theta$ a predicted response waveform $h(\theta)\in\mathcal{H}$.

We assume the mapping satisfies standard regularity and identifiability conditions:
\begin{enumerate}
\item[(i)] \emph{Smoothness}: $h\in C^2(\Theta,\mathcal{H})$, so that first and second partial derivatives with respect to coordinates exist and are continuous in the Hilbert-space norm.
\item[(ii)] \emph{Local injectivity}: the differential $\dd h_\theta:T_\theta\Theta\to\mathcal{H}$ has full rank $m$ in the domain of interest, ensuring that the tangent vectors $t_i(\theta):=\partial_i h(\theta)$ are linearly independent:
\begin{equation}
\sum_i v^i\,\partial_i h(\theta)=0\quad\Longrightarrow\quad v=0.
\end{equation}
\item[(iii)] \emph{Positive-definite Fisher metric}: the induced Gram matrix
\begin{equation}
g_{ij}(\theta)=\ip{\partial_i h(\theta)}{\partial_j h(\theta)}
\end{equation}
is finite and positive definite on the identifiable parameter subspace.
\end{enumerate}

In physical waveform families, exact or approximate degeneracies can occur due to
parameter symmetries or weak sensitivities.
Geometrically, this corresponds to $\dd h_\theta$ failing to be injective, or equivalently the
Fisher matrix $g_{ij}$ becoming singular
or ill-conditioned. In such cases we restrict attention to an \emph{identifiable subspace}
$\mathcal{V}_\theta\subset T_\theta\Theta$ on which $g_{ij}$ is positive definite, e.g.\ by removing
degenerate directions or using a pseudoinverse with a controlled singular-value threshold.
All projection formulas below are understood on this identifiable subspace. The limits of
quadratic (Fisher) approximations in the presence of degeneracies are discussed in~\cite{Vallisneri2008}.

\subsection{Fisher metric from the likelihood}\label{subsec:fisher}
The log-likelihood (up to an additive constant independent of $\theta$) is
\begin{equation}
\ln L(\theta;d)=-\frac12\norm{d-h(\theta)}^2+\text{const},
\label{eq:loglik}
\end{equation}
so that, for stationary Gaussian noise, maximum-likelihood estimation is
equivalent to \emph{minimizing the squared distance} between the data and the
model waveform in the noise-weighted Hilbert space. With the residual
$r(\theta):=d-h(\theta)$, which equals the noise $n$ at the true
parameters, and $\partial_i r(\theta)=-\partial_i h(\theta)$,
differentiating the quadratic form gives the score
\begin{equation}
\partial_i\ln L(\theta;d)=-\frac12\,\partial_i\braket{r(\theta)|r(\theta)}
=-\braket{r(\theta)|\partial_i r(\theta)}
=\braket{r(\theta)|\partial_i h(\theta)},
\label{eq:score}
\end{equation}
the projection of the residual onto the tangent directions $\partial_i h$.
At the maximum-likelihood point $\hat\theta$ these projections vanish,
\begin{equation}
\braket{d-h(\hat\theta)|\partial_i h(\hat\theta)}=0\quad\forall i,
\label{eq:mle_orthogonality}
\end{equation}
i.e.\ the optimal residual is orthogonal to the tangent space of the
waveform manifold. Differentiating once more,
\begin{equation}
\partial_i\partial_j\ln L(\theta;d)
=-\braket{\partial_j h(\theta)|\partial_i h(\theta)}
+\braket{r(\theta)|\partial_i\partial_j h(\theta)}.
\label{eq:hessian}
\end{equation}
Taking the expectation over the noise at the true parameters, where
$r=n$ and $\E[n]=0$, removes the second term and yields the Fisher information matrix
as the negative expected Hessian,
\begin{equation}
g_{ij}(\theta):=-\E\!\left[\partial_i\partial_j\ln L(\theta;d)\right]
=\braket{\partial_i h(\theta)|\partial_j h(\theta)},
\label{eq:fisher}
\end{equation}
the Fisher--Rao metric on parameter space. In the high-SNR regime the random
curvature term $\braket{n|\partial_i\partial_j h}$ is subleading, so
$g_{ij}$ controls the local quadratic approximation to the likelihood and
defines the induced Riemannian metric on parameter space.

\subsection{Embedded manifold}
The image $\mathcal{M}_1:=h(\Theta)\subset\mathcal{H}$ constitutes the single-source waveform manifold. Wherever $\dd h_\theta$ has full rank, $h$ is an immersion and $\mathcal{M}_1$ is locally an embedded submanifold of dimension $m=\dim\Theta$; we assume in addition that $h$ is injective on the domain of interest, so that $\mathcal{M}_1$ has no self-intersections there and is embedded (an immersion alone would not exclude them):
\begin{equation}
\mathcal{M}_1=\{h(\theta)\in\mathcal{H}\mid \theta\in\Theta\}.
\end{equation}

The ambient inner product in $\mathcal{H}$ induces a Riemannian metric on $\mathcal{M}_1$ via the Fisher matrix. Tangent vectors at $h(\theta)$ take the form
\begin{equation}
v=v^i\,\partial_i h(\theta),\qquad w=w^j\,\partial_j h(\theta),
\end{equation}
with inner product
\begin{equation}
\ip{v}{w}=v^i w^j\,\ip{\partial_i h(\theta)}{\partial_j h(\theta)}=v^i w^j\,g_{ij}(\theta).
\end{equation}

Waveform similarity is conventionally quantified by the noise-weighted \emph{overlap}, or match,
$\langle \hat h(\theta)\mid \hat h(\theta')\rangle\in[-1,1]$ between normalized waveforms
$\hat h(\theta):=h(\theta)/\|h(\theta)\|$, and the corresponding \emph{mismatch}
\begin{equation}
\mu(\theta,\theta'):=1-\langle \hat h(\theta)\mid \hat h(\theta')\rangle.
\end{equation}
For nearby parameters $\theta'=\theta+\Delta\theta$, Taylor expansion gives
\begin{equation}
\mu(\theta,\theta+\Delta\theta)
= \frac12\,\bar g_{ij}(\theta)\,\Delta\theta^i\Delta\theta^j+\mathcal{O}(\|\Delta\theta\|^3),
\end{equation}
where the local mismatch metric is
\begin{equation}
\bar g_{ij}(\theta):=\langle \partial_i \hat h(\theta)\mid \partial_j \hat h(\theta)\rangle.
\end{equation}
This Riemannian structure underpins template-bank placement~\cite{Owen1996,OwenSathyaprakash1999} and local
mismatch control~\cite{CutlerFlanagan1994,Owen1996}.
Here, by contrast, we focus on \emph{extrinsic} deviations orthogonal to the
single-source manifold that cannot be absorbed by shifting tangent parameters.

Fix $\theta\in\Theta$ and denote the tangent vectors in $\mathcal{H}$ by
\begin{equation}
t_i(\theta):=\partial_i h(\theta)\in\mathcal{H}.
\end{equation}
The induced Gram matrix is $g_{ij}(\theta)=\langle t_i\mid t_j\rangle$.
Assuming $g_{ij}$ is invertible on the identifiable subspace, define $g^{ij}$ by
$g^{ik}g_{kj}=\delta^i{}_j$.
Then the orthogonal projection of an arbitrary $v\in\mathcal{H}$ onto the tangent space is
\begin{equation}
P_T[v]
:= t_i\,g^{ij}\,\langle t_j\mid v\rangle,
\end{equation}
and the normal projection is~\cite{doCarmo1992,Lee2018Riemannian}
\begin{equation}
P_N[v]:=v-P_T[v].
\end{equation}
Equivalently, the normal space is
\begin{equation}
N_{h(\theta)}\mathcal{M}_1=\{v\in\mathcal{H}:\langle v\mid t_i\rangle=0\ \forall i\}.
\end{equation}
If $h\in C^k$ and $g_{ij}$ remains uniformly positive definite on a compact region of
parameter space, then $P_T(\theta)$ and $P_N(\theta)$ vary smoothly with $\theta$ (at least
$C^{k-1}$), ensuring controlled Taylor expansions under parameter changes~\cite{Lee2018Riemannian}.

\subsection{Intrinsic versus extrinsic geometry; second fundamental form}\label{subsec:II}
Because $\mathcal{M}_1\subset\mathcal{H}$ is embedded in a (flat) Hilbert space, second
derivatives in $\mathcal{H}$ admit a canonical decomposition into tangent and normal parts.
In coordinates, define
\begin{equation}
h_{ij}(\theta):=\partial_i\partial_j h(\theta)\in\mathcal{H}.
\end{equation}
Then
\begin{equation}
h_{ij}=P_T[h_{ij}] + P_N[h_{ij}]
= \Gamma^{k}{}_{ij}\,t_k + \mathrm{II}_{ij},
\end{equation}
where the Christoffel symbols of the Levi--Civita connection induced by $g_{ij}$ are
\begin{equation}
\Gamma^{k}{}_{ij}
= g^{k\ell}\,\langle t_\ell\mid h_{ij}\rangle,
\end{equation}
and the \emph{second fundamental form} (an $\mathcal{H}$-valued symmetric bilinear form) is
\begin{equation}
\mathrm{II}_{ij}:=\mathrm{II}(\partial_i,\partial_j):=P_N[h_{ij}].
\end{equation}
The second fundamental form $\mathrm{II}$ isolates the acceleration of the
waveform mapping orthogonal to the tangent space. As derived in
Section~\ref{sec:local}, the norm of the normal curvature vector
$\mathrm{II}(u,u)$ along unit parameter directions $u$ governs the
leading-order distinguishability between nearby overlapping
sources~\cite{Lee2018Riemannian}.

\section{Multi-source waveforms, hypotheses, and a distance-to-manifold statistic}\label{sec:multi}

On the Cartesian product $\Theta^N=\Theta\times\cdots\times\Theta$, the set of $N$-source superpositions defines the subset
\begin{equation}
\mathcal{M}_N:=\left\{\sum_{k=1}^N h(\theta_k)\ \Big|\ \theta_k\in\Theta\right\}\subset\mathcal{H}.
\end{equation}
For two sources, $\mathcal{M}_2=\{h(\theta_1)+h(\theta_2)\mid \theta_1,\theta_2\in\Theta\}$.
The assumption that the composite detector strain is an additive superposition,
\begin{equation}
h_I(t)=\sum_{k=1}^{N} h_I(\theta_k)(t),
\end{equation}
rests on two physical properties. First, metric perturbations obey the linearized vacuum Einstein equations in the radiation zone, so independent perturbations superpose at the detector (any strong-field nonlinearities near the sources being encapsulated in the individual models $h(\theta_k)$). Second, the detector response---whether antenna patterns in ground-based interferometers or time-delay interferometry (TDI) combinations in LISA---acts as a linear functional of incident strain~\cite{TintoDhurandhar2014,ColpiEtAl2024LISARedBook}.
For ground-based detectors with antenna patterns $F_{+,\times}(\Omega,\psi)$,
\begin{equation}
h(t)=F_{+}(\Omega,\psi)\,h_{+}(t)+F_{\times}(\Omega,\psi)\,h_{\times}(t),
\end{equation}
and for LISA the TDI observables are linear combinations of delayed inter-spacecraft phase measurements.
Permutation symmetry among identical sources renders the parameterization $\Theta^N\to\mathcal{M}_N$ many-to-one, and $\mathcal{M}_N$ is generally not an embedded manifold globally (possessing self-intersections when parameters coincide); nonetheless, it forms a well-defined subset of the ambient Hilbert space $\mathcal{H}$ in which likelihood comparisons are evaluated.

The multi-source problem is especially prominent for LISA: in the millihertz band
($10^{-4}\,\mathrm{Hz}\lesssim f\lesssim 1\,\mathrm{Hz}$), tens of thousands of
Galactic compact binaries will overlap continuously alongside merging massive black-hole binaries
and extreme-mass-ratio inspirals~\cite{JaranowskiKrolak2005,AmaroSeoaneEtAl2017,TimpanoRubboCornish2006,TintoDhurandhar2014,ColpiEtAl2024LISARedBook,Maggiore2008,ESAFactsheet2024,AmaroSeoaneEtAl2023LRR,RobsonCornishLiu2019,BaghiEtAl2022LDC,BandopadhyayMoore2024Yorsh,LISADataChallengeWebsite}.
The LISA Data Challenges construct simulated TDI data from such superpositions, and global-fit pipelines recover blended populations from these streams~\cite{StrubEtAl2024,BaghiEtAl2022LDC}.
These analyses use frequency-domain Gaussian likelihoods with channel-dependent noise spectral densities, inducing the noise-weighted inner product on the direct sum of TDI channels~\cite{ColpiEtAl2024LISARedBook,RobsonCornishLiu2019}. The distance-to-manifold statistic $D(d)=\inf_{\theta}\|d-h(\theta)\|$ is defined in that same inner product, directly governing the maximized single-source likelihood. Frequency- and sky-dependent responses beyond the long-wavelength approximation are incorporated directly into the waveform mapping $h(\theta)$~\cite{TintoDhurandhar2014,ValenciaHusa2025}.

\subsection{Model selection as a geometric problem}
Consider model comparison between a single-source hypothesis $H_1$ and a two-source hypothesis $H_2$:
\begin{align}
H_1:&\quad d(t)=h(\theta)(t)+n(t),\quad d-n\in\mathcal{M}_1,\\
H_2:&\quad d(t)=h(\theta_1)(t)+h(\theta_2)(t)+n(t),\quad d-n\in\mathcal{M}_2.
\end{align}
Under $H_2$ the log-likelihood is
\begin{equation}
\ln L_2(\theta_1,\theta_2;d)=-\frac12\norm{d-h(\theta_1)-h(\theta_2)}^2+\text{const}.
\end{equation}
Discriminating between $H_1$ and $H_2$ is therefore geometrically equivalent to evaluating the proximity of data $d$ to the sets $\mathcal{M}_1$ and $\mathcal{M}_2$ in the noise-weighted geometry of $\mathcal{H}$.

We define the noise-weighted distance from an arbitrary waveform $\omega\in\mathcal{H}$ to the single-source manifold $\mathcal{M}_1$ as
\begin{equation}
D(\omega):=\inf_{\theta\in\Theta}\norm{\omega-h(\theta)}=\inf_{\theta\in\Theta}\sqrt{\ip{\omega-h(\theta)}{\omega-h(\theta)}},
\end{equation}
and its square
\begin{equation}
D^2(\omega):=\inf_{\theta\in\Theta}\norm{\omega-h(\theta)}^2.
\end{equation}
Operationally, $D^2(\omega)$ is the smallest squared residual achievable when approximating $\omega$ by a single source.
Equivalently, $-\tfrac12 D^2(\omega)$ is, up to an additive constant, the maximized log-likelihood under $H_1$:
\begin{equation}
\max_{\theta}\ln L_1(\theta;\omega)=-\frac12 D^2(\omega)+\text{const}.
\end{equation}

When $\Theta$ is compact and $h$ is continuous, $\mathcal{M}_1=h(\Theta)$ is compact and the infimum is attained as a minimum:
\begin{equation}
\theta_{\mathrm{eff}}(\omega)\in\arg\min_{\theta\in\Theta}\norm{\omega-h(\theta)}.
\end{equation}
The corresponding effective waveform is $h_{\mathrm{eff}}(\omega):=h(\theta_{\mathrm{eff}}(\omega))\in\mathcal{M}_1$, with residual $\omega-h_{\mathrm{eff}}(\omega)$. At any interior minimum, first-order optimality requires this residual to be orthogonal to the tangent space $T_{h_{\mathrm{eff}}}\mathcal{M}_1$, a property central to the perturbative expansion in Section~\ref{sec:local}.

\subsection{Likelihood-ratio interpretation}\label{subsec:lr}
To isolate the geometric structure of the model-selection problem independently of noise fluctuations, consider first the noiseless case $d=\omega\in\mathcal{M}_2$.
Under $H_1$, maximizing the likelihood over $\theta$ minimizes the squared distance to $\mathcal{M}_1$:
\begin{equation}
\ln L^{\max}_1(\omega)=\max_{\theta\in\Theta}\ln L_1(\theta;\omega)
=-\frac12\min_{\theta\in\Theta}\norm{\omega-h(\theta)}^2+\text{const}
=-\frac12 D^2(\omega)+\text{const}.
\end{equation}
Under $H_2$, evaluating the likelihood at the true parameters $\theta_1,\theta_2$ gives an exact match, $\ln L^{\max}_2(\omega)=\text{const}$.
The maximized log-likelihood ratio between the hypotheses is therefore
\begin{equation}
\Delta\ln L_{\max}(\omega):=\ln L^{\max}_2(\omega)-\ln L^{\max}_1(\omega)
=\frac12 D^2(\omega).
\end{equation}
In the noiseless baseline, the squared distance $D^2(\omega)$ from the two-source waveform to the single-source manifold directly governs the preference for $H_2$ over $H_1$.

Two sources are effectively indistinguishable from a single template whenever their residual falls below an operational threshold $D_*>0$, defining the confusion region
\begin{equation}
\mathcal{R}_2(D_*):=\{\omega\in\mathcal{M}_2\mid D(\omega)\le D_*\}.
\end{equation}
A decision threshold $\Delta_0$ on the maximized log-likelihood ratio corresponds to
\begin{equation}
\Delta_0=\frac12 D_*^2\quad\Longleftrightarrow\quad D_*=\sqrt{2\Delta_0}.
\end{equation}
In realistic analyses the data contain noise, $d=\omega+n$. Model selection
between the two nested hypotheses is then governed by the maximized log-likelihood ratio
\begin{equation}
\Lambda_{21}(d):=\ln L^{\max}_2(d)-\ln L^{\max}_1(d)
=\frac12\left[D_1^2(d)-D_2^2(d)\right],
\qquad
D_k^2(d):=\inf_{\omega'\in\mathcal{M}_k}\norm{d-\omega'}^2,
\label{eq:glrt}
\end{equation}
which reduces to $\tfrac12 D^2(\omega)=\Delta\ln L_{\max}(\omega)$ in the
noiseless baseline, where $D_2^2(\omega)=0$. In the presence of noise, the
distribution of this likelihood ratio requires separate calibration.
At coincident source parameters, the two-source model is singular even when
both amplitudes are positive: source separation is unidentifiable to first
order. Consequently, a standard $\chi^2$ calibration and an identification of
$D^2(\omega)$ as its noncentrality parameter do not follow from the local
geometric expansion alone.

Throughout this work, $D_*$ is therefore adopted as an operational geometric
tolerance. Assigning false-alarm and detection probabilities requires an
appropriate analysis of the nonregular likelihood ratio~\cite{Chernoff1954,SelfLiang1987,Drton2009}
or calibration with noise injections using the same parameter constraints,
noise model, and maximization procedure as the intended analysis (Section~\ref{sec:conclusions}).
Within the local quartic regime, changing $D_*^2$ rescales curvature-limited
separation estimates in proportion to $(D_*^2)^{1/4}$; this geometric scaling does
not itself provide a statistical calibration.

\section{Local extrinsic geometry and perturbative resolvability}\label{sec:local}

The most difficult regime for source distinguishability arises when two signals have nearly identical parameters. Consider a reference source $\theta_0\in\Theta$ and a secondary source at $\theta_0+\delta\theta$ with small displacement $\delta\theta$, generating the noiseless two-source waveform
\begin{equation}
\omega(\delta\theta)=h(\theta_0)+h(\theta_0+\delta\theta).
\label{eq:omega_local}
\end{equation}
Throughout this section we assume:
\begin{enumerate}
\item[(A)] \textbf{Regularity.} $h:\Theta\to\mathcal{H}$ is $C^3$ as a map into $\mathcal{H}$, so that Taylor expansions through second order have a controlled $\mathcal{O}(\|\delta\theta\|^3)$ remainder in $\mathcal{H}$-norm.
\item[(B)] \textbf{Identifiability.} The Fisher matrix $g_{ij}$ is positive definite on the identifiable subspace in a neighbourhood of $\theta_0$ (Section~\ref{sec:single}).
\item[(C)] \textbf{Local uniqueness.} For $\omega$ near the single-source waveform that fits the coincident pair exactly ($2h(\theta_0)$ for equal amplitudes, the point $\gamma_0$ of Section~\ref{subsec:ift}), the minimizer of $\norm{\omega-h(\theta)}$ is locally unique on the identifiable subspace, so that $\theta_{\mathrm{eff}}(\omega)$ is well defined there; along an exact degeneracy only the fitted waveform, not the parameter vector, is unique.
\end{enumerate}
Physical waveform families satisfy these conditions away from singular boundaries. Post-Newtonian and phenomenological approximants depend smoothly on intrinsic parameters, the only potential non-smooth features being abrupt frequency cut-offs, which our numerical model in Section~\ref{sec:numerical-validation} replaces with smooth tapers.
When exact parameter degeneracies arise---such as the single spin combination governing 1.5PN aligned-spin phasing---the Fisher matrix acquires an exact null space. In such cases, projection operators are restricted to the identifiable subspace, and Section~\ref{sec:numerical-validation} demonstrates how these flat directions manifest in the confusion geometry.
Local uniqueness cannot be guaranteed globally on a nonlinear manifold: the branch guaranteed by the implicit function theorem (Section~\ref{subsec:ift}) represents the global nearest point only within a neighbourhood where no distant sheet of $\mathcal{M}_1$ approaches closer. In Section~\ref{subsec:numerical-recipes} we verify this numerically by repeating every optimization from an independent starting point; while not an analytical guarantee of global optimality, this provides practical evidence against competing local minima.

From the first-order condition (\ref{eq:mle_orthogonality}), the optimal residual $r_{\mathrm{eff}}(\omega):=\omega-h(\theta_{\mathrm{eff}}(\omega))$ is orthogonal to the tangent space at the best-fit point:
\begin{equation}
P_T(\theta_{\mathrm{eff}})\,r_{\mathrm{eff}}=0,\qquad r_{\mathrm{eff}}=P_N(\theta_{\mathrm{eff}})\,r_{\mathrm{eff}}.
\end{equation}
The normal bundle $N\mathcal{M}_1$ therefore governs the leading irreducible mismatch~\cite{Lee2018Riemannian}.

\subsection{Amplitude--shape split}\label{subsec:ampshape}
We decompose the waveform map into an overall amplitude and normalized shape:
\begin{equation}
h(\theta)=A\,s(\nu),\qquad \theta=(A,\nu),
\label{eq:ampshape}
\end{equation}
where $A>0$ absorbs the luminosity distance, together with any orientation parameter that acts as a common scale on every channel and harmonic (with several harmonics and channels the inclination and polarization change the shape as well and then belong to $\nu$), and $\nu$ collects the shape parameters. In coordinates $\gamma=(A_{\mathrm{eff}},\nu)$, the tangent vectors are $t_0=s(\nu)$ and $t_i=A_{\mathrm{eff}}\,\partial_i s(\nu)$. The tangent space $\mathrm{span}\{s,\partial_i s\}$ and projectors $P_T, P_N$ are therefore invariant under changes of $A_{\mathrm{eff}}$ along a ray of fixed shape $\nu$. Writing $s:=s(\nu)$, $s_i:=\partial_i s(\nu)$, $s_{ij}:=\partial_i\partial_j s(\nu)$, and
\begin{equation}
b_{ij}:=P_N[s_{ij}],
\label{eq:b_def}
\end{equation}
the second fundamental form at $A_{\mathrm{eff}}\,s(\nu)$ is
\begin{equation}
\mathrm{II}(\partial_i,\partial_j)=A_{\mathrm{eff}}\,b_{ij},
\qquad
\mathrm{II}(\partial_A,\,\cdot\,)=0,
\label{eq:II_vs_b}
\end{equation}
since $\partial_A\partial_i h=s_i$ is tangent and $\partial_A^2 h=0$. The amplitude direction is flat, and $\mathrm{II}$ scales linearly with the amplitude at the point of evaluation.

\subsection{Existence and uniqueness of the local best fit}\label{subsec:ift}
For equal amplitudes, $\theta_0=(A,\nu)$ and
$\theta_0+\delta\theta=(A,\nu+\delta\theta)$, a pure shape displacement, the exactly degenerate
configuration $\delta\theta=0$ is fitted by the single source
$\gamma_0=(2A,\nu)$, $h(\gamma_0)=2A\,s(\nu)$. We look for the best fit near
$\gamma_0$ as the root of the stationarity system
\begin{equation}
I_\alpha(\gamma,\delta\theta):=
\ip{\omega(\delta\theta)-h(\gamma)}{\partial_\alpha h(\gamma)}=0,
\qquad \alpha=0,\ldots,m-1,
\label{eq:stationarity_system}
\end{equation}
the first-order condition of the squared-distance functional
$F(\gamma;\delta\theta)=\norm{\omega(\delta\theta)-h(\gamma)}^2$; although
$h$ takes values in $\mathcal{H}$, the stationarity equations are scalar
inner products, so $I$ is a finite-dimensional system. Since
$\omega(0)=h(\gamma_0)$ we have $I(\gamma_0,0)=0$, and
\begin{equation}
\left.\pd{I_\alpha}{\gamma^\beta}\right|_{(\gamma_0,0)}
=-\ip{\partial_\beta h(\gamma_0)}{\partial_\alpha h(\gamma_0)}
+\ip{\omega(0)-h(\gamma_0)}{\partial_\alpha\partial_\beta h(\gamma_0)}
=-g_{\alpha\beta}(\gamma_0).
\label{eq:ift_jacobian}
\end{equation}
By (B) the Jacobian is invertible on the identifiable subspace, so the
implicit function theorem~\cite{Lee2013ISM} provides a neighbourhood of
$\delta\theta=0$ and a unique smooth map
$\gamma^*(\delta\theta)=\gamma_0+\Delta\gamma(\delta\theta)$ with
$I(\gamma^*(\delta\theta),\delta\theta)=0$; this is
$\theta_{\mathrm{eff}}(\omega(\delta\theta))$ of Section~\ref{sec:multi} in
the present coordinates, and its smoothness justifies the order-by-order
expansion below. The theorem provides a local minimum branch; $D^2$ below
is the distance to this branch, which is the distance to $\mathcal{M}_1$
whenever no other part of the manifold lies closer, as assumed in (C). With
the residual
$r(\gamma;\delta\theta):=\omega(\delta\theta)-h(\gamma)$ the distance is
\begin{equation}
D^2(\omega(\delta\theta))
=\norm{r(\gamma^*(\delta\theta);\delta\theta)}^2
=\norm{P_N\,r(\gamma^*(\delta\theta);\delta\theta)}^2 .
\label{eq:distance_normal}
\end{equation}

\subsection{Perturbative calculation: equal amplitudes}\label{subsec:equal}
Expanding $s(\nu+\delta\theta)$ to second order gives
\begin{equation}
\omega(\delta\theta)=2A\,s+A\,s_i\,\delta\theta^i
+\frac{A}{2}\,s_{ij}\,\delta\theta^i\delta\theta^j
+\mathcal{O}(\|\delta\theta\|^3),
\label{eq:omega_expansion}
\end{equation}
and the effective source is parameterized as
$h_{\mathrm{eff}}(\alpha,\varepsilon)=(2A+\alpha)\,s(\nu+\varepsilon)$ with a
small amplitude correction $\alpha$ and a small shape shift $\varepsilon$.
Solving the stationarity conditions order by order in $\delta\theta$
(Appendix~\ref{app:equal}) gives, at first order,
\begin{equation}
\varepsilon^{(1)i}=\tfrac12\,\delta\theta^i,\qquad \alpha^{(1)}=0:
\label{eq:first_order_equal}
\end{equation}
the effective source sits at the midpoint of the pair and the first-order
residual, which lies in the tangent span $\{s,s_i\}$, is cancelled
completely. At second order the tangent part of the residual is again
cancelled by suitable (and, under (B), unique) $\alpha^{(2)}$,
$\varepsilon^{(2)}$, while its normal part is irreducible:
\begin{equation}
r_N(\delta\theta)=\frac{A}{4}\,b_{ij}\,\delta\theta^i\delta\theta^j
+\mathcal{O}(\|\delta\theta\|^3)\quad\text{in }\mathcal{H}.
\label{eq:rN_equal}
\end{equation}
Under (A) the remainder is controlled in $\mathcal{H}$-norm, so squaring
gives
\begin{equation}
D^2(\omega(\delta\theta))=\norm{r_N(\delta\theta)}^2
=\frac{A^2}{16}\,C_{ijk\ell}\,
\delta\theta^i\delta\theta^j\delta\theta^k\delta\theta^\ell
+\mathcal{O}(\|\delta\theta\|^5),
\qquad C_{ijk\ell}:=\ip{b_{ij}}{b_{k\ell}},
\label{eq:quartic_tensor}
\end{equation}
the cross term between the quadratic residual and the cubic remainder being
$\mathcal{O}(\|\delta\theta\|^5)$; $C$ is symmetric under $i\leftrightarrow j$,
$k\leftrightarrow\ell$ and $(ij)\leftrightarrow(k\ell)$.

To state the result in a coordinate-free form we measure the separation with
the Fisher metric at the base point,
$\delta^2:=g_{ij}(\theta_0)\,\delta\theta^i\delta\theta^j$, and write
$\delta\theta^i=\delta\,u^i$ with $g_{ij}u^iu^j=1$. Then
\begin{equation}
D^2(\delta,u)=\frac{A^2}{16}\,K(u)\,\delta^4+\mathcal{O}(\delta^5),
\qquad
K(u):=C_{ijk\ell}\,u^iu^ju^ku^\ell=\norm{b(u,u)}^2 .
\label{eq:quartic_law}
\end{equation}

\paragraph{Geometric interpretation.}
By (\ref{eq:II_vs_b}), $A\,b_{ij}$ is the second fundamental form of
$\mathcal{M}_1\hookrightarrow\mathcal{H}$ at the source point
$h(\theta_0)=A\,s(\nu)$, so the law reads
\begin{equation}
D^2(\delta,u)=\frac{1}{16}\,
\norm{\mathrm{II}_{h(\theta_0)}(u,u)}^2\,\delta^4+\mathcal{O}(\delta^5):
\label{eq:quartic_law_II}
\end{equation}
the leading irreducible mismatch is one sixteenth of the squared norm of the
normal curvature vector at the source along the displacement direction. Here $\mathrm{II}$ is an \emph{extrinsic} geometric quantity describing how the manifold bends within ambient waveform space, distinct from the intrinsic curvature of parameter space~\cite{doCarmo1992,Lee2018Riemannian}.
Evaluating $\mathrm{II}$ at the effective point $2A\,s(\nu)$ rather than at the source point $A\,s(\nu)$ scales the second fundamental form by a factor of two, yielding the equivalent form $\tfrac{1}{64}\norm{\mathrm{II}_{2As}(u,u)}^2\delta^4$.
Directions of small normal curvature correspond to weakly curved embeddings that permit severe degeneracies; conversely, directions of large extrinsic curvature generate substantial normal residuals, rendering the sources distinguishable at smaller separations.

\subsection{Operational distinguishability scale}\label{subsec:dist}
Imposing a tolerance $D_*$ on the distance statistic,
$D^2(\delta,u)\gtrsim D_*^2$, and keeping the leading term gives the
directional distinguishability scale
\begin{equation}
\delta_{\min}(u)\simeq\left(\frac{16D_*^2}{A^2K(u)}\right)^{1/4}.
\label{eq:delta_min}
\end{equation}
This estimate applies when $K(u)>0$ and the predicted separation lies
within the perturbative regime. At fixed signal strength and tolerance,
smaller $K(u)$ then requires larger separation for distinguishability.
Within this regime, the separation estimate scales in proportion to
$(D_*^2)^{1/4}$. When $K(u)=0$, higher-order terms must be examined unless
an exact waveform symmetry establishes degeneracy.

\subsection{Unequal amplitudes}\label{subsec:unequal}
For sources $\theta_1=(A_1,\nu)$ and $\theta_2=(A_2,\nu+\delta\theta)$,
the relevant case for LISA, where distances and orientations differ, the
two-source waveform is
$\omega=(A_1+A_2)\,s+A_2\,s_i\delta\theta^i
+\tfrac{A_2}{2}s_{ij}\delta\theta^i\delta\theta^j+\mathcal{O}(\|\delta\theta\|^3)$,
and the effective source is sought near $(A_\Sigma,\nu)$,
$A_\Sigma:=A_1+A_2$. The same order-by-order minimization
(Appendix~\ref{app:unequal}) gives $\alpha^{(1)}=0$ and
\begin{equation}
\varepsilon^{(1)i}=\frac{A_2}{A_\Sigma}\,\delta\theta^i,
\qquad
r^{(2)}_N(\delta\theta)=\frac{A_1A_2}{2(A_1+A_2)}\,b_{ij}\,
\delta\theta^i\delta\theta^j:
\label{eq:unequal_results}
\end{equation}
the effective source sits at the amplitude-weighted centroid of the pair,
and the irreducible residual carries the coefficient $A_{\mathrm{harm}}/4$
with the harmonic-mean amplitude
$A_{\mathrm{harm}}:=2A_1A_2/(A_1+A_2)$. The quartic law therefore keeps its
form with $A\mapsto A_{\mathrm{harm}}$,
\begin{equation}
D^2(\delta,u)=\frac{A_{\mathrm{harm}}^2}{16}\,K(u)\,\delta^4
+\mathcal{O}(\delta^5),
\qquad
\delta_{\min}(u)\simeq
\left(\frac{16D_*^2}{A_{\mathrm{harm}}^2K(u)}\right)^{1/4}.
\label{eq:quartic_law_unequal}
\end{equation}
For $q=A_2/A_1$ the prefactor relative to the equal-amplitude law at
amplitude $A_1$ is $(2q/(1+q))^2$, the quantity tested numerically in
Section~\ref{subsec:quartic-scaling-tests}.

\section{Numerical methods and results}
\label{sec:numerical-validation}

To test the perturbative scaling predictions against direct likelihood minimization, we implement an inspiral waveform model incorporating the orbital motion of the LISA constellation, subdominant harmonics, and instrumental plus Galactic confusion noise. We evaluate the distance-to-manifold statistic across representative parameter directions, verify the predicted quartic scaling, and map the resulting confusion zones.
All computations are carried out with the open-source Julia package \textsc{CurvatureDistinguishability.jl}~\cite{CurvatureDistinguishability2026}, which implements the waveform and detector response, the geometric curvature diagnostic, and the constrained optimization routines described below.

The single-source model is a six-parameter frequency-domain inspiral with parameter vector
\begin{equation}
\theta=(D_L,\mathcal{M},t_c,\phi_c,\chi_1,\chi_2),
\label{eq:numerical_state_vector}
\end{equation}
comprising the luminosity distance, detector-frame chirp mass, coalescence time, reference phase, and two aligned dimensionless spins. Waveforms are evaluated in the noise-weighted Hilbert space of Section~\ref{sec:noise} through the LISA response, ensuring that the inner product used in the theoretical definition of $D(\omega)$ matches the one evaluated numerically for residuals, projections, and curvature. In our numerical realization, the amplitude scale is absorbed directly into the whitened response $h^w$ (Section~\ref{subsec:geometric-diagnostic-implementation}). For a direction $u$ normalized by the Fisher metric at reference source $\theta_0$,
\begin{equation}
D^2_{\mathrm{th}}(\delta,u)=\frac{1}{16}\,K_{\mathrm{num}}(u)\,\delta^4,
\qquad
K_{\mathrm{num}}(u):=\big\|P_N\big[\partial^2_u h^w(\theta_0)\big]\big\|^2
=\norm{\mathrm{II}_{h(\theta_0)}(u,u)}^2=A^2K(u),
\label{eq:numerical_quartic_law}
\end{equation}
reproducing (\ref{eq:quartic_law_II}); for an amplitude ratio $q=A_2/A_1$, the right-hand side acquires the factor $(2q/(1+q))^2$ from (\ref{eq:quartic_law_unequal}).

Separations are measured throughout in Fisher units. For a direction $u$ with $g_{ij}u^iu^j=1$, the separation $\delta=1$, one Fisher $\sigma$, is the displacement $\delta u$ at which the linearized change of the waveform, $\delta\,u^i\partial_i h$, has unit norm; it is the point at which the line through $\theta_0$ along $u$ crosses the unit Fisher ellipsoid. This is the conditional scale of the direction, with all other parameters held fixed, not a marginalized Cram\'er--Rao uncertainty; its size in physical units varies by orders of magnitude between directions (Section~\ref{subsec:quartic-scaling-tests}).

Given a tolerance $D_*$, the local distinguishability boundary along $u$ is $\delta_{\min}(u)=(16D_*^2/K_{\mathrm{num}}(u))^{1/4}$. Throughout this section, $D_*^2=1$ (corresponding to a residual SNR of unity) provides the reference geometric unit; as discussed in Section~\ref{subsec:lr}, any other tolerance simply rescales every boundary by $(D_*^2)^{1/4}$.

We employ two complementary numerical tests:
\begin{enumerate}
\item[(i)] \emph{One-dimensional parameter sweeps}: Fixing a reference source $\theta_0$ and displacement direction $u$, we place a second source at $\theta_0+\delta u$ across thirty log-spaced separations $\delta$ from $10^{-3}$ to $10^{2.5}\,\sigma$. For each separation, the two-source waveform $\omega_\delta=h(\theta_0)+h(\theta_0+\delta u)$ is fitted with a single template by minimizing over the physically admissible box $\Theta_{\mathrm{phys}}$ ($D_L,\mathcal{M},t_c\ge 0$, $|\chi_{1,2}|\le 1$, and periodic phase):
\begin{equation}
D^2_{\mathrm{opt}}(\delta)=
\min_{\theta_{\mathrm{eff}}\in\Theta_{\mathrm{phys}}}
\norm{\omega_\delta-h(\theta_{\mathrm{eff}})}^2,
\label{eq:numerical_distance_minimization}
\end{equation}
directly testing the theoretical law (\ref{eq:numerical_quartic_law}).
\item[(ii)] \emph{Two-dimensional confusion maps}: The directional curvature $K_{\mathrm{num}}(u)$ is evaluated around the circle of directions in selected parameter planes and inverted into $\delta_{\min}(u)$. The enclosed region, intersected with physical priors, delineates the local confusion zone.
\end{enumerate}

\subsection{Waveform, detector response and noise model}
\label{subsec:numerical-waveform-detector}

Rather than a full production approximant, the model is an inspiral description
designed to isolate the physical mechanisms governing local degeneracy breaking:
chirp-mass phase evolution, coalescence-time and phase shifts, aligned spin--orbit
coupling, subdominant harmonics, the finite observation window, and the orbital
modulation of the detector. Internally
every parameter is rescaled to order unity, a conditioning step for the
Fisher matrix, the automatic-differentiation derivatives and the constrained
Newton iteration: $D_L$ is measured in Gpc, $\mathcal{M}$ in geometrized
seconds ($1\,\mathrm{s}\simeq 2.03\times10^{5}\,M_\odot$) and $t_c$ in units
of $10^{6}\,\mathrm{s}$, with $\phi_c$ and $\chi_{1,2}$ dimensionless.
Distances and projections are nevertheless computed in the noise-weighted
waveform space, so the rescaling is only a coordinate choice.

The symmetric mass ratio is a fixed property of the scenario,
$\eta=0.2222$ (mass ratio $2\!:\!1$ to four significant figures, the value used in every computation), with total mass
$M=\mathcal{M}\eta^{-3/5}$ and mass asymmetry
$\delta_m=(m_1-m_2)/M=\sqrt{1-4\eta}\simeq1/3$; the heavier body carries
$\chi_1$. In the stationary-phase approximation the harmonic $k$ of the
orbital phase radiates at frequency $f$ when the dominant $(2,2)$ harmonic
radiates at $F=2f/k$. With $v=(\pi MF)^{1/3}$ the post-Newtonian phase of
the dominant harmonic is the TaylorF2 phase to 1.5PN
order~\cite{Poisson1995,Arun2009},
\begin{equation}
\psi_{\mathrm{PN}}(F)=\frac{3}{128\,\eta\,v^{5}}
\left[1+\left(\frac{3715}{756}+\frac{55}{9}\eta\right)v^{2}
+\left(4\beta-16\pi\right)v^{3}\right],
\label{eq:psi_pn_numerical_model}
\end{equation}
with the spin--orbit coefficient
\begin{equation}
\beta=\frac{1}{12}\left[(113-76\eta)\,\chi_s+113\,\delta_m\,\chi_a\right],
\qquad
\chi_s=\tfrac12(\chi_1+\chi_2),\quad \chi_a=\tfrac12(\chi_1-\chi_2).
\label{eq:beta_numerical_model}
\end{equation}
The spins enter the phase only through the scalar $\beta$: at this order
exactly one spin combination is a flat direction of the signal manifold at
every mass ratio ($\chi_a$ at equal mass, in general the line
$(113-76\eta+113\delta_m)\,\mathrm{d}\chi_1+(113-76\eta-113\delta_m)\,\mathrm{d}\chi_2=0$,
$\mathrm{d}\chi_2/\mathrm{d}\chi_1=-2.29$ at $\eta=0.2222$), so the tangent
space has rank five rather than six and the Fisher matrix is singular. This
is a property of the 1.5PN aligned-spin phasing, not of the numerical
resolution; the first term that lifts it is the 2PN spin--spin coupling.
The tangent-space construction of
Section~\ref{subsec:geometric-diagnostic-implementation} detects this
direction automatically, and the spin--spin confusion map of
Section~\ref{subsec:confusion-zone-maps} displays it. The phase of
harmonic $k$ is
\begin{equation}
\Psi_k(f)=2\pi f t_c-\frac{k}{2}\phi_c+\frac{k}{2}\,\psi_{\mathrm{PN}}(2f/k)-\frac{\pi}{4},
\qquad \tilde h_k(f)\propto e^{-\ii\Psi_k(f)},
\label{eq:psi_k_numerical_model}
\end{equation}
and its stationary-phase emission time is the Newtonian
$t_k(f)=t(2f/k)$ with
\begin{equation}
t(F)=t_c-\frac{5\mathcal{M}}{256\,(\pi\mathcal{M}F)^{8/3}},
\label{eq:time_frequency_relation_numerical}
\end{equation}
consistent with $\partial_f\Psi_k=2\pi t_k$ at leading order.

The polarizations carry the amplitude corrections of
Blanchet, Iyer, Will and Wiseman to 0.5PN order~\cite{Blanchet1996},
\begin{equation}
h_{+,\times}=\frac{2M\eta\,x}{D_L}
\left[H^{(0)}_{+,\times}+x^{1/2}H^{(1/2)}_{+,\times}\right],
\qquad x=(\pi MF)^{2/3},
\label{eq:biww_amplitudes}
\end{equation}
with $H^{(0)}_+=-(1+c_\iota^2)\cos2\varphi$, $H^{(0)}_\times=-2c_\iota\sin2\varphi$,
$H^{(1/2)}_+=-\tfrac{s_\iota}{8}\delta_m[(5+c_\iota^2)\cos\varphi-9(1+c_\iota^2)\cos3\varphi]$
and $H^{(1/2)}_\times=-\tfrac34 s_\iota\,c_\iota\,\delta_m[\sin\varphi-3\sin3\varphi]$,
where $c_\iota=\cos\iota$, $s_\iota=\sin\iota$ and $\varphi$ is the orbital phase.
The stationary-phase amplitude of harmonic $k$ is
$a_k(f)=(M\eta x/D_L)\sqrt{2/(k\dot F)}$ with
$\dot F=\tfrac{96}{5}\pi^{8/3}\mathcal{M}^{5/3}F^{11/3}$ evaluated at
$F=2f/k$; for $k=2$ this reproduces the standard
$\sqrt{5/24}\,\pi^{-2/3}\mathcal{M}^{5/6}f^{-7/6}(1+c_\iota^2)/(2D_L)$, while the
$k=1,3$ harmonics are proportional to $\delta_m$. The subdominant harmonics
are thus physical, with amplitudes fixed by the mass ratio, eliminating the
artificial degeneracies of a quadrupole-only model. Every harmonic
is multiplied by a smooth window at the Schwarzschild innermost stable
circular orbit, $f_{\mathrm{ISCO}}=(6^{3/2}\pi M)^{-1}$,
\begin{equation}
w_k(f)=\tfrac12\left[1-\tanh\frac{F/f_{\mathrm{ISCO}}-1}{\sigma_{\mathrm{ISCO}}}\right],
\qquad F=2f/k,\quad \sigma_{\mathrm{ISCO}}=0.1,
\label{eq:isco_window}
\end{equation}
which ends the inspiral without a hard cut-off, keeping every quantity a
smooth function of $\theta$ and $f$ as the geometric construction requires;
merger and ringdown are not modelled.

The frequency-domain signal contains every frequency of the band regardless
of when it was emitted. For the lighter systems this matters: at the lower band edge
the harmonic $k=2$ was emitted $0.13\,\mathrm{yr}$ before coalescence for
$\mathcal{M}=4\,\mathrm{s}$, but $14\,\mathrm{yr}$ for
$\mathcal{M}=0.25\,\mathrm{s}$ and $63\,\mathrm{yr}$ for
$\mathcal{M}=0.1\,\mathrm{s}$, long before a one-year observation starts.
Each harmonic is therefore weighted by the window of an observation over
$[0,T]$ evaluated at its emission time,
\begin{equation}
w_k(f)\;\to\;w_k(f)\,W\!\big(t_k(f)\big),
\qquad
W(t)=\tfrac12\left[\tanh\frac{t}{\Delta}-\tanh\frac{t-T}{\Delta}\right],
\label{eq:observation_window}
\end{equation}
with an edge time scale $\Delta=10^{6}\,\mathrm{s}$ (a few times the local
chirp time $\dot f^{-1/2}$ at $t=0$ for the lightest systems, so that the
stationary-phase treatment of the window holds). The window is smooth in
$\mathcal{M}$ and $t_c$, through which its edge moves in frequency; it
removes the band below $F_0$ with $t(F_0)=0$ and leaves the signal above
about $2F_0$ unchanged. For the heavy base points below, the window is
inert; for the two light ones it changes the confusion geometry along the
mass and spin directions, and the corresponding maps are computed with it.

The constellation follows the analytic equal-arm orbits of Rubbo, Cornish
and Poujade~\cite{Rubbo2004} to second order in the eccentricity, at an
orbital radius of $1\,\mathrm{AU}$ and with the orbital phase $\alpha(t)=2\pi t/\mathrm{yr}$:
the constellation plane is inclined by $60^\circ$ to the ecliptic and
cartwheels once per year. A wavefront from the source direction $\hat n$
reaches the constellation centre $\vec R_c(t)$ before the solar-system
barycentre, so the barycentric strain acquires the Doppler phase
$\Delta_{\mathrm D}(f,t)=2\pi f\,\hat n\cdot\vec R_c(t)$, evaluated at the
emission time $t_k(f)$ of each harmonic. This coupling is essential for the
coalescence-time direction: shifting $t_c$ relocates the detector along its
orbit at every contributing frequency. In the long-wavelength limit each
spacecraft synthesizes a $60^\circ$ Michelson interferometer with the detector
tensor $D_i=\tfrac12(\hat a_i\otimes\hat a_i-\hat b_i\otimes\hat b_i)$ from
its two arm directions, $F^{+,\times}_i=D_i:e^{+,\times}$, with the
polarization tensors built in the ecliptic frame of the LISA science ground
segment conventions~\cite{Baghi2026SGSConventions}. The three Michelson responses $X,Y,Z$ are
combined into the noise-orthogonal channels of Prince et al.~\cite{Prince2002}, $A=(Z-X)/\sqrt2$, $E=(X-2Y+Z)/\sqrt6$; since
$X+Y+Z$ vanishes at leading order in $f/f_\star$ for noise and signal alike,
the channels are rescaled by $\sqrt{2/3}$ so that each carries the
single-Michelson noise spectral density, $F_A=(F_Z-F_X)/\sqrt3$ and
$F_E=(F_X-2F_Y+F_Z)/3$, and the null combination $T$ is identically zero and
excluded. The observed strain of channel $C\in\{A,E\}$ is
\begin{equation}
\tilde h_C(f)=\sum_{k=1}^{3}w_k(f)\,\mathcal{T}(f)
\left[a^{+}_k(f)F^{+}_C(t_k)-\ii\,a^{\times}_k(f)F^{\times}_C(t_k)\right]
e^{-\ii\Psi_k(f)}\,e^{+\ii\Delta_{\mathrm D}(f,t_k)},
\label{eq:channel_strain_numerical_model}
\end{equation}
where $\mathcal{T}(f)=[1+0.6(f/f_\star)^2]^{-1/2}$ is the finite-arm
transfer roll-off of Robson, Cornish and Liu~\cite{RobsonCornishLiu2019}
(their Eq.~9) with $f_\star=c/(2\pi L_{\mathrm{arm}})=19.09\,\mathrm{mHz}$ for
the arm length $L_{\mathrm{arm}}=2.5\times10^{9}\,\mathrm{m}$. The numerical inner product sums the two
channels,
\begin{equation}
\braket{a|b}_{\mathrm{num}}=4\,\Re\sum_{I\in\{A,E\}}
\int_{f_{\min}}^{f_{\max}}\frac{\tilde a_I^*(f)\tilde b_I(f)}{S_I(f)}\,\dd f,
\label{eq:numerical_channel_inner_product}
\end{equation}
discretized on the frequency grid below.

The noise weighting follows the analytic LISA model of Robson, Cornish and
Liu~\cite{RobsonCornishLiu2019}. The instrumental part is their
Michelson-channel PSD (their Eq.~12),
\begin{equation}
P_n(f)=\frac{P_{\mathrm{OMS}}}{L_{\mathrm{arm}}^2}
+2\left(1+\cos^2(f/f_\star)\right)\frac{P_{\mathrm{acc}}}{(2\pi f)^4L_{\mathrm{arm}}^2},
\label{eq:numerical_instrumental_psd}
\end{equation}
with the optical-metrology and test-mass acceleration contributions
$P_{\mathrm{OMS}}$ and $P_{\mathrm{acc}}$ of their Eqs.~10--11. $P_n$ is the
noise of one channel and is used as such: their sky- and
polarization-averaged response $R(f)=\tfrac{3}{10}\,[1+0.6(f/f_\star)^2]^{-1}$
(their Eq.~9) is not applied to it, because the response is modelled
explicitly. The unresolved Galactic foreground enters through their
analytic confusion fit (their Eq.~14),
\begin{equation}
S_c(f)=A_c f^{-7/3}e^{-f^{\alpha}+\beta f\sin(\kappa f)}
\left[1+\tanh\!\big(\gamma(f_k-f)\big)\right],
\qquad A_c=9\times10^{-45},
\label{eq:numerical_confusion_psd}
\end{equation}
with the one-year coefficients of their Table~1 ($\alpha=0.171$,
$\beta=292$, $\kappa=1020$, $\gamma=1680$, $f_k=2.15\,\mathrm{mHz}$). This
fit is calibrated on the two-channel sensitivity curve, $S_n=P_n/R+S_c$
(their Eq.~1), so it enters the per-channel PSD through the response,
\begin{equation}
S_I(f)=P_n(f)+R(f)\,S_c(f),\qquad I\in\{A,E\},
\label{eq:numerical_channel_psd}
\end{equation}
and dividing $S_I$ by $R$ recovers their sensitivity curve, so that the
sky- and polarization-averaged signal-to-noise ratio of the $A+E$
combination coincides with the one computed from that curve. After the
conversion the Galactic term still exceeds $P_n$ by an order of magnitude
at $1\,\mathrm{mHz}$: the confusion bump dominates the noise budget in the
millihertz band and directly reweights the inner products that define
$K_{\mathrm{num}}(u)$.

The grid is a one-year observation, $T=3.15576\times10^{7}\,\mathrm{s}$
(the orbital period entering the constellation motion), over
$f\in[10^{-4},\,5\times10^{-2}]\,\mathrm{Hz}$ at spacing
$\Delta f=1/T$: $1.57\times10^{6}$ frequency bins. This
resolution is needed because the residuals of the quartic law can lie many
orders of magnitude below the signal power, and it resolves the beating of
the harmonics and of nearby sources. Source direction (ecliptic longitude
$180^\circ$, latitude $30^\circ$), inclination $30^\circ$ and polarization
angle $45^\circ$ are common to all cases. Table~\ref{tab:base_points} lists
the six reference sources of the campaign; they are loud
($\rho\simeq 7\times10^{2}$--$1.7\times10^{3}$ in $A+E$), so one Fisher
$\sigma$ is a relative change of $3\times10^{-7}$--$9\times10^{-6}$ in the
chirp mass and the distinguishability scales $\delta_{\min}$ are tens to over a
hundred $\sigma$.

\begin{table}[t]
\caption{Reference sources of the campaign (detector-frame masses;
distances of 2--8\,Gpc correspond to $z\simeq0.35$--$1.2$; every coalescence
lies inside the one-year observation; $\phi_c$ in radians). $\rho$ is the single-source
signal-to-noise ratio in the $A+E$ channels. The sweeps and maps below are
labelled by these base points.}
\label{tab:base_points}
\centering
\footnotesize
\setlength{\tabcolsep}{4pt}
\begin{tabular}{llccccccccc}
\hline
& Description & $D_L$ [Gpc] & $\mathcal{M}$ [$M_\odot$] & $M$ [$M_\odot$] & $t_c$ [d] & $\phi_c$ & $\chi_1$ & $\chi_2$ & $f_{\mathrm{ISCO}}$ [mHz] & $\rho$ \\
\hline
B1 & distant, moderate mass & 8 & $3.2\times10^{5}$ & $8.0\times10^{5}$ & 174 & 1.57 & 0.6 & $-0.4$ & 5.5 & 1734 \\
B2 & massive, aligned spins & 5 & $8.1\times10^{5}$ & $2.0\times10^{6}$ & 231 & 0 & 0.8 & 0.8 & 2.2 & 694 \\
B2$'$ & massive, anti-aligned spins & 5 & $8.1\times10^{5}$ & $2.0\times10^{6}$ & 231 & 0 & 0.3 & $-0.3$ & 2.2 & 694 \\
B3 & strongly anti-aligned spins & 5 & $5.1\times10^{5}$ & $1.3\times10^{6}$ & 116 & 1.57 & 0.7 & $-0.7$ & 3.5 & 1195 \\
B4 & low mass, band-filling & 3 & $5.1\times10^{4}$ & $1.3\times10^{5}$ & 289 & 0 & 0.9 & 0 & 35 & 1361 \\
B5 & light seed & 2 & $2.0\times10^{4}$ & $5.0\times10^{4}$ & 231 & 0 & 0.7 & 0.5 & 88 & 762 \\
\hline
\end{tabular}
\end{table}

\subsection{Geometric diagnostic and its numerical realization}
\label{subsec:geometric-diagnostic-implementation}

The numerical diagnostic reproduces the geometric construction of the
analytic derivation. For each channel $I$ and frequency bin $f_k$ the
whitened waveform components are
\begin{equation}
h^w_{I,k}(\theta)=2\sqrt{\frac{\Delta f}{S_I(f_k)}}\,\tilde h_I(f_k;\theta),
\label{eq:whitened_waveform_components}
\end{equation}
so that the Euclidean norm of the whitened vector reproduces the discretized
noise-weighted inner product (\ref{eq:numerical_channel_inner_product}).
All distances, residuals and projections are computed in the
finite-dimensional real Hilbert space obtained by stacking the real and
imaginary parts of the whitened channels, the numerical counterpart of
$\mathcal{H}$.

For a reference configuration the code first constructs the tangent space
of the single-source manifold from the parameter derivatives
$t_\alpha=\partial h^w/\partial\gamma^\alpha$ at the base point. The
numerical Fisher matrix is their Gram matrix,
$g^{\mathrm{num}}_{\alpha\beta}=\braket{t_\alpha|t_\beta}_{\mathrm{num}}$.
For the repeated projections the tangent vectors are orthonormalized by a
modified Gram--Schmidt procedure in the noise-weighted inner product;
directions whose orthogonalized norm falls below a fixed tolerance are
exactly (or numerically) degenerate and are excluded from the basis. For
the present model this happens at every base point: the 1.5PN null spin
combination of Section~\ref{subsec:numerical-waveform-detector} is
detected automatically, the basis has rank five, and all projections are
taken on the identifiable subspace, since inverting the Gram matrix would be
ill-conditioned in exactly the configurations of interest.
With the orthonormal basis $e_a$, $P_Tv=\sum_a e_a\braket{e_a|v}_{\mathrm{num}}$
and $P_Nv=v-P_Tv$.

All waveform derivatives are computed by forward-mode automatic
differentiation with dual numbers, exact to machine precision. The tangent
vectors come from one Jacobian evaluation of the whitened response map; for
a direction $u$ the map $\epsilon\mapsto h^w(\theta_0+\epsilon u)$ is
evaluated once with nested dual numbers, which yields the value, the first
directional derivative and the second directional derivative
$H(u)=\partial^2_\epsilon h^w(\theta_0+\epsilon u)|_{\epsilon=0}
=u^\alpha u^\beta\,\partial_\alpha\partial_\beta h^w$ from the same
evaluation. The directional curvature is then
\begin{equation}
K_{\mathrm{num}}(u)=\big\|P_N H(u)\big\|^2_{\mathrm{num}}
=\big\|H(u)\big\|^2_{\mathrm{num}}
-\sum_a\big|\braket{e_a|H(u)}_{\mathrm{num}}\big|^2 :
\label{eq:numerical_directional_curvature}
\end{equation}
every component of the directional Hessian that can be represented by
motion along the single-source manifold is subtracted, and only the normal
component remains. $H(u)$, and hence $K_{\mathrm{num}}(u)$, is exactly even
under $u\to-u$; the maps exploit this symmetry. The boundary point in a
parameter plane is independent of the normalization of the direction
vector (rescaling $u\to\lambda u$ multiplies $K_{\mathrm{num}}$ by
$\lambda^4$ and $\delta_{\min}$ by $\lambda^{-1}$), so the curvature can
be evaluated along raw coordinate directions with no intermediate Fisher
rescaling entering the boundary.

\subsection{Numerical recipes and safeguards}
\label{subsec:numerical-recipes}

Because the numerical sweeps probe residuals spanning nineteen orders of magnitude in $D^2$, specific safeguards are required to ensure robust convergence and prevent optimizer artifacts from biasing the scaling tests.

\paragraph{Optimization.} The best-fit single source of
(\ref{eq:numerical_distance_minimization}) is found with an interior-point
Newton method using the exact automatic-differentiation gradient and Hessian
of the objective, subject to the physical box constraints, initialized near
the degenerate effective source ($(2A,\nu_0)$ for equal amplitudes,
$((1+q)A,\nu_0+\tfrac{q}{1+q}\delta u)$ for an amplitude ratio $q$).
Convergence is declared on the projected gradient norm or on a relative
change of $D^2$ below $10^{-8}$ between Newton steps; iteration counts,
gradient norms and the activity of every bound are recorded for each fit.
No physical bound is active at any optimum of the campaign, so the
constraint never distorts the scaling test. Every fit is repeated from a
seeded, randomly perturbed second starting point: across the 210
separations of the seven sweeps the second start never found a lower
minimum above the optimizer floor beyond tolerance-level refinements of the
same minimum (relative gains below $3\times10^{-5}$); order-unity gains occur
only at floor points, where the objective is flat to round-off, so no
competing basin was found above the floor; this is numerical evidence for
assumption (C) of Section~\ref{sec:local}, not a proof of uniqueness or of
global optimality.

\paragraph{Optimizer floor and fit windows.} The smallest separations of
every sweep reach the precision floor of the optimized objective, the
double-precision resolution of the fitted parameters, at
$D^2\simeq 10^{-16}$--$10^{-13}$. The floor is detected automatically as the
contiguous run of smallest separations whose ratio
$D^2_{\mathrm{opt}}/D^2_{\mathrm{th}}$ exceeds a fixed threshold, and every
point at or below the detected level is excluded from all fits and marked in
the figures. Two quantities are fitted to the clean points. The higher-order
coefficients of the ratio, $D^2_{\mathrm{opt}}/D^2_{\mathrm{th}}\simeq
1+c_1\delta+c_2\delta^2$, use the points within $30\,\%$ of the leading-order
law (the perturbative window); they describe the departure inside the
window and are not extrapolated beyond the last clean point, where a
two-term fit has no predictive value. The range over which the law holds
is reported from the measured ratio itself, as the largest separation of
the grid at which the ratio stays within a stated fraction of unity. The
exponent of the law is fitted over the same perturbative window; a fit over
all clean points is recorded as well and is quoted where the two differ,
since separations beyond the end of the perturbative regime (Section~\ref{subsec:quartic-scaling-tests}) are not described by the law they would
otherwise be fitted to. The quoted uncertainties of the exponents are the
least-squares standard errors of the log--log regression over deterministic
points: they measure the scatter of the points about a power law, not a
statistical error, and the fitted values are effective slopes over the
stated windows.

\paragraph{Cross-architecture execution.} The benchmark campaign shipped
with the package, seven sweeps of thirty separations and seven maps at 8192
mirrored directions on the $1.57\times10^{6}$-bin grid in the configuration
of Section~\ref{subsec:numerical-waveform-detector} without the observation
window, was executed on the CPUs of six hosts of two vendors and six
processor generations
(Table~\ref{tab:runtime_cpu}); the results are bit-identical on five of
them and agree to $5\times10^{-6}$ in $D^2$ on the sixth, the Ryzen~9 9950X
(a seventh host with a second 9950X reproduced the first bit for bit and is
not tabulated), so that the arithmetic of the implementation does not
depend on the host. The results reported in
Sections~\ref{subsec:quartic-scaling-tests}--\ref{subsec:residual-spectra}
were produced in a single run of the production configuration, with the
window, on the Core Ultra~7 155H; the timing tables and the agreement
statistics of this paragraph refer to the benchmark campaign. On the two
hybrid Intel processors, which combine performance
and efficiency cores of different throughput, the Julia threads were
confined to the performance cores (sixteen threads pinned to the eight
P-cores of the i9-13900KS through the scheduler affinity mask, twelve
threads, the P-core thread count, on the Core Ultra~7 155H), so that each
timing refers to a homogeneous set of cores; the i7-10750H and the AMD
processors have homogeneous cores and used every hardware thread. The sweep
stage, which is dominated by loss, gradient and Hessian evaluations of the
$1.57\times10^{6}$-bin objective, was in addition re-executed through the
GPU kernel path on six accelerators of three vendors (a data-centre
H200~NVL, consumer, mobile and workstation cards, and an integrated GPU) at every
setting of the Hessian chunk on the discrete cards and at one setting on the
integrated GPU (Table~\ref{tab:runtime_gpu}). On the GPU the Hessian is evaluated in
chunks of dual-number lanes (\texttt{hessian\_chunk} in the configuration):
the setting $c=0$ evaluates the full $6\times6$ Hessian in one launch of
49 lanes, $c=3$, $2$ and $1$ in $4$, $9$ and $36$ launches of $16$, $9$ and
$4$ lanes. Every completed GPU execution reproduces the CPU reference: the
optimized distances agree to $3\times10^{-4}$ along the near-null spin
direction and to $2\times10^{-4}$ or better along every other direction, at
every chunk setting, and the fitted exponents agree within their
uncertainties. The setting changes the wall time of a sweep by up to a factor of
$2.2$ on the same card: on the CUDA devices the 9- and 16-lane kernels are
the fastest and the 4-lane setting the slowest, the untiled 49-lane kernel
in between although it performs the least arithmetic; on the two RDNA~3
cards the 49-lane kernel is the fastest. The
missing entries of Table~\ref{tab:runtime_gpu} are of two kinds. On the
ROCm devices individual runs stopped making progress with the device idle
and no error reported, a lost dispatch completion in the ROCm~7.1
runtime that we could not resolve, more frequent on the W7900 at the
settings with more launches per Hessian, and were terminated by the
per-sweep timeout; the affected sweeps were not repeated. On the integrated
GPU the 49-lane kernel exceeds the device's limit on the size of kernel
arguments and does not compile, and a 16-lane launch runs within a factor
of two of the display driver's preemption timeout, so only $c=2$ was run
to completion. A separate two-year-grid timing benchmark of the same fit,
computed but not shown here, reproduces the ordering of the chunk settings
on every accelerator.

\begin{table}[t]
\caption{Wall time (h:mm) of the benchmark campaign on six CPUs: the
seven sweeps S1--S7 (each thirty separations with two optimizer starts on
the $1.57\times10^{6}$-bin grid), the map stage (seven planes at 8192
mirrored directions) and the whole run. Thread counts in the header: every
hardware thread on the homogeneous-core processors, the performance-core
threads on the two hybrid Intel parts. Sweeps
(Section~\ref{subsec:quartic-scaling-tests}): S1 B1 diagonal, S2 B2 chirp mass and
coalescence time, S3 the same with $q=1/2$, S4 B3 phase and spins, S5 B4
coalescence time, S6 B5 chirp mass and spins, S7 B2$'$ near-null spin
direction.}
\label{tab:runtime_cpu}
\centering
\scriptsize
\setlength{\tabcolsep}{4pt}
\begin{tabular}{lcccccc}
\hline
& Ryzen 9 9950X & i9-13900KS & Ryzen 9 5950X & EPYC 7551P & Ultra 7 155H & i7-10750H \\
Threads & 32 & 16 & 32 & 64 & 12 & 12 \\
\hline
S1 & 1:36 & 2:24 & 2:36 & 4:06 & 4:15 & 6:27 \\
S2 & 1:10 & 1:38 & 1:47 & 2:48 & 3:01 & 4:04 \\
S3 & 1:28 & 2:00 & 2:25 & 3:51 & 3:58 & 5:28 \\
S4 & 1:35 & 1:59 & 2:40 & 4:12 & 4:16 & 6:03 \\
S5 & 1:41 & 1:53 & 2:30 & 3:55 & 3:32 & 5:33 \\
S6 & 1:25 & 2:10 & 2:36 & 3:57 & 3:47 & 5:19 \\
S7 & 1:57 & 2:25 & 3:08 & 5:14 & 4:49 & 6:09 \\
Map stage & 1:29 & 2:06 & 2:33 & 2:51 & 4:26 & 5:31 \\
\hline
Total & 12:20 & 16:35 & 20:15 & 30:55 & 32:03 & 44:33 \\
\hline
\end{tabular}
\end{table}

\begin{table}[t]
\caption{Wall time (h:mm) of the seven sweeps re-executed through the GPU
kernel path on six accelerators, at every Hessian chunk setting $c$ the
device supports: $(1+c)^2$ dual-number lanes per launch and
$\lceil 6/c\rceil^{2}$ launches per Hessian, $c=0$ being the full 49-lane
kernel. The H200 settings $c=0,3$ and $c=1,2$ ran on two cards of the same
node. $\dag$: the run stopped making progress on the ROCm runtime and was
terminated by the per-sweep timeout; the total is given for complete rows
only. On the integrated GPU $c=0$ does not compile and $c=1,3$ were not
run (Section~\ref{subsec:numerical-recipes}). The host CPU is not repeated
here; the map stage runs on the host.}
\label{tab:runtime_gpu}
\centering
\scriptsize
\setlength{\tabcolsep}{3.5pt}
\begin{tabular}{llcccccccc}
\hline
Device & $c$ & S1 & S2 & S3 & S4 & S5 & S6 & S7 & Total \\
\hline
H200 NVL (CUDA) & 0 & 0:13 & 0:11 & 0:11 & 0:13 & 0:14 & 0:17 & 0:15 & 1:35 \\
 & 1 & 0:16 & 0:14 & 0:14 & 0:17 & 0:16 & 0:20 & 0:16 & 1:52 \\
 & 2 & 0:11 & 0:09 & 0:09 & 0:11 & 0:11 & 0:14 & 0:11 & 1:16 \\
 & 3 & 0:11 & 0:09 & 0:10 & 0:11 & 0:11 & 0:14 & 0:13 & 1:20 \\
\hline
RTX 5090 (CUDA) & 0 & 0:23 & 0:20 & 0:19 & 0:24 & 0:25 & 0:32 & 0:28 & 2:50 \\
 & 1 & 0:31 & 0:27 & 0:26 & 0:33 & 0:32 & 0:42 & 0:32 & 3:43 \\
 & 2 & 0:18 & 0:16 & 0:15 & 0:19 & 0:18 & 0:24 & 0:19 & 2:10 \\
 & 3 & 0:17 & 0:14 & 0:15 & 0:18 & 0:17 & 0:23 & 0:20 & 2:04 \\
\hline
RTX 2080 S, mobile (CUDA) & 0 & 2:27 & 2:07 & 2:01 & 2:38 & 2:44 & 3:25 & 2:44 & 18:06 \\
 & 1 & 2:53 & 2:28 & 2:23 & 3:03 & 3:14 & 3:54 & 3:00 & 20:55 \\
 & 2 & 1:38 & 1:27 & 1:24 & 1:47 & 1:52 & 2:20 & 1:50 & 12:18 \\
 & 3 & 1:47 & 1:34 & 1:35 & 1:54 & 1:56 & 2:31 & 2:04 & 13:20 \\
\hline
W7900 (ROCm) & 0 & 0:31 & 0:23 & 0:27 & 0:31 & 0:29 & 0:30 & \dag & --- \\
 & 1 & \dag & \dag & 0:51 & \dag & \dag & \dag & \dag & --- \\
 & 2 & \dag & \dag & 0:43 & \dag & \dag & \dag & \dag & --- \\
 & 3 & \dag & 0:24 & 0:29 & 0:33 & 0:31 & 0:33 & 0:34 & --- \\
\hline
RX 7700 XT (ROCm) & 0 & 0:51 & 0:35 & 0:39 & \dag & 0:45 & 0:50 & 0:54 & --- \\
 & 1 & 1:44 & 1:15 & 1:21 & 1:37 & 1:39 & 1:53 & \dag & --- \\
 & 2 & 1:01 & 0:42 & 0:47 & 0:55 & 0:55 & 1:02 & 1:04 & 6:27 \\
 & 3 & 0:53 & 0:37 & 0:41 & 0:47 & \dag & 0:54 & 0:57 & --- \\
\hline
Arc iGPU (oneAPI) & 2 & 6:08 & 4:10 & 5:01 & 5:13 & 5:17 & 5:52 & 6:16 & 37:57 \\
\hline
\end{tabular}
\end{table}

\subsection{Quartic-law validation across parameter-space directions}
\label{subsec:quartic-scaling-tests}

\begin{figure}[tbp]
    \centering
    \includegraphics[width=\textwidth]{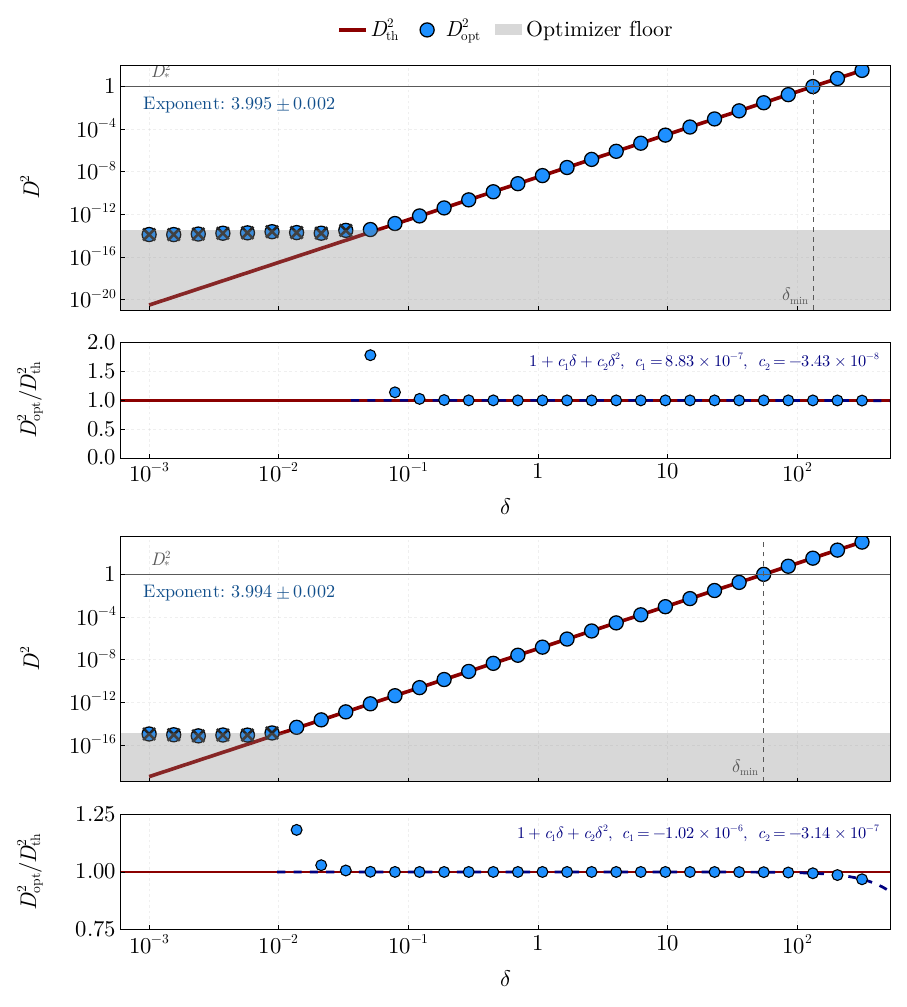}
    \caption{
    Quartic-law validation along two well-measured directions. Each panel
    shows the optimized squared distance $D^2_{\mathrm{opt}}$ (markers)
    against the prediction $D^2_{\mathrm{th}}=\tfrac{1}{16}K_{\mathrm{num}}(u)\delta^4$
    (line), the residual threshold $D_*^2=1$ with its $\delta_{\min}$, the
    optimizer-floor band (points inside it struck through and excluded from
    every fit), and, in the strip beneath its main axis, the ratio
    $D^2_{\mathrm{opt}}/D^2_{\mathrm{th}}$ with the fitted correction
    $1+c_1\delta+c_2\delta^2$. The annotated exponent is fitted over the
    perturbative window. Upper panel: B1, five-parameter diagonal
    direction $u\propto(0,0.707,-0.5,0.3,0.4,-0.1)$, exponent
    $3.995\pm0.002$. Lower panel: B2, chirp-mass/coalescence-time direction
    $u\propto(0,1,0.5,0,0,0)$, $3.994\pm0.002$.
    }
    \label{fig:scaling_diagonal_masstime}
\end{figure}

\begin{figure}[tbp]
    \centering
    \includegraphics[width=\textwidth]{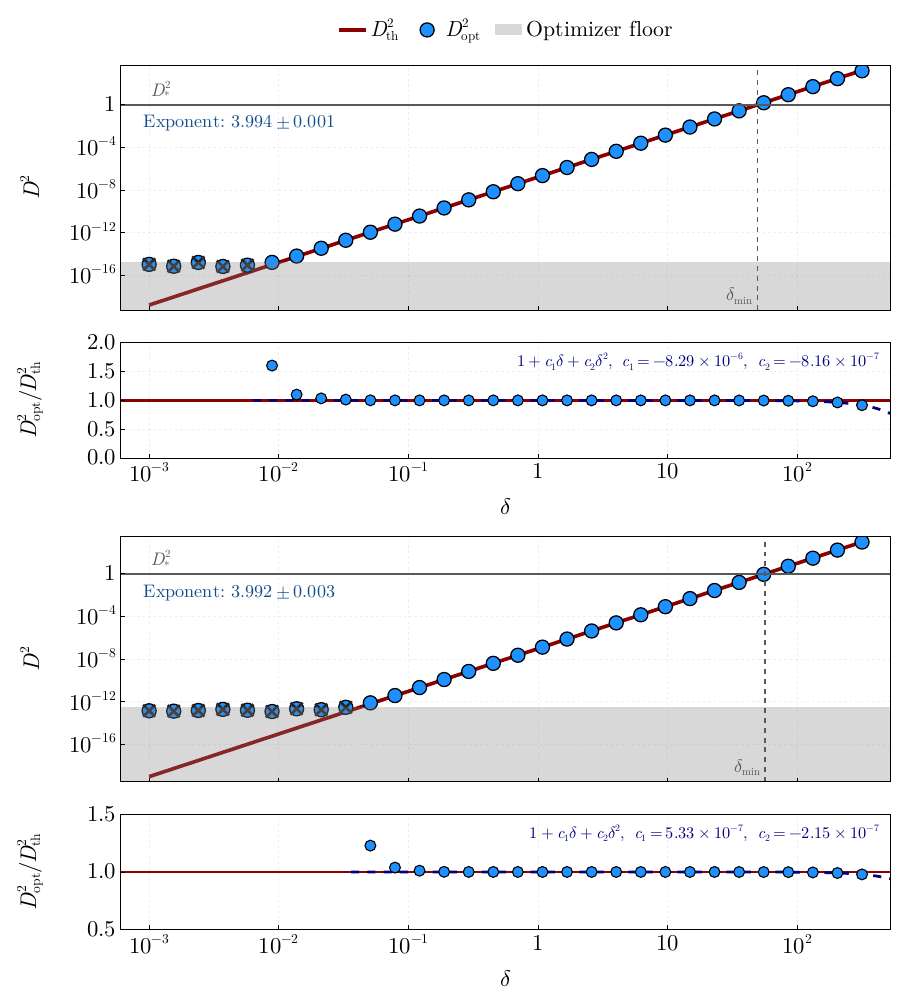}
    \caption{
    Quartic-law validation in the spin sector and along the detector-motion
    direction, in the layout of Figure~\ref{fig:scaling_diagonal_masstime}.
    Upper panel: B3, phase and both spins, $u\propto(0,0,0,1,0.5,0.5)$,
    exponent $3.994\pm0.001$. Lower panel: B4, pure coalescence-time shift
    $u=e_{t_c}$, $3.992\pm0.003$.
    }
    \label{fig:scaling_spin_time}
\end{figure}

\begin{figure}[tbp]
    \centering
    \includegraphics[width=\textwidth]{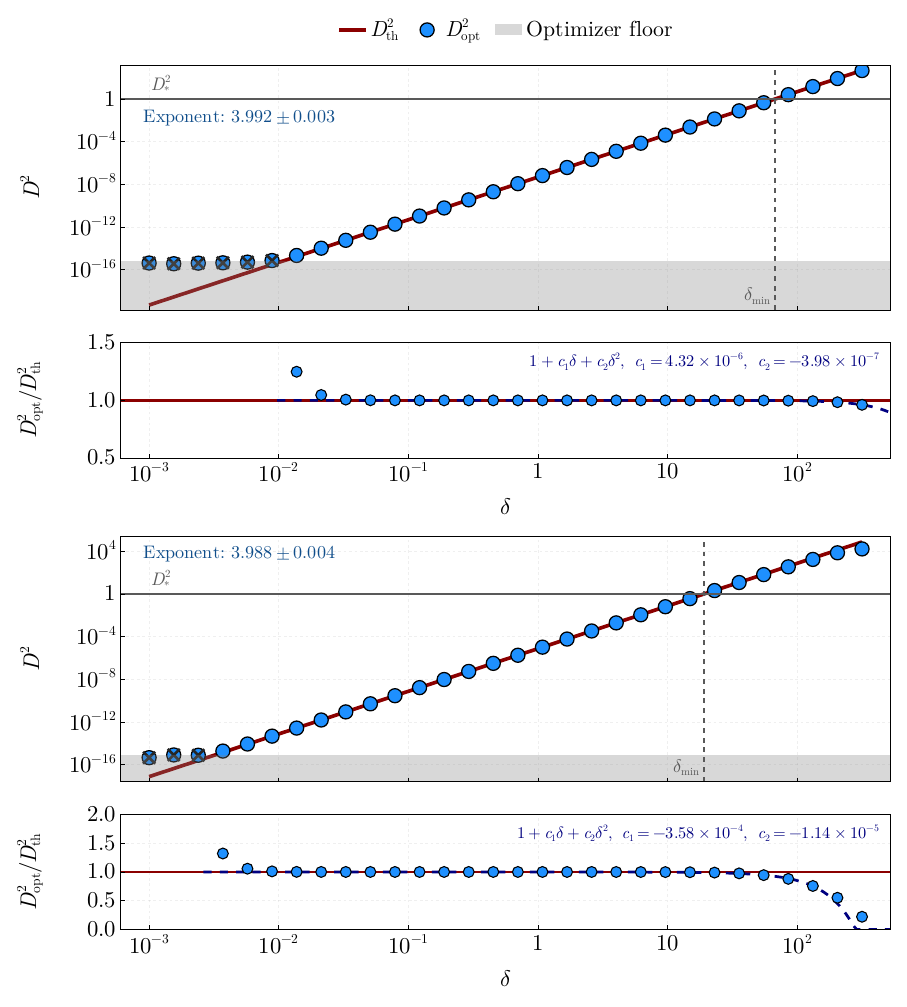}
    \caption{
    Prefactor test and validity boundary, in the layout of
    Figure~\ref{fig:scaling_diagonal_masstime}. Upper panel: B2 along the
    mass--time direction with an amplitude ratio $q=A_2/A_1=1/2$; the
    prediction carries the harmonic-mean factor $(2q/(1+q))^2=4/9$ of
    (\ref{eq:quartic_law_unequal}); exponent $3.992\pm0.003$, and at matched
    separations the measured $D^2(q{=}1/2)/D^2(q{=}1)$ lies between
    $0.44442$ and $0.44448$ against $4/9=0.44444$. Lower panel: B2$'$ along
    $u\propto(0,0.15,0,0,0.4,-0.92)$, a near-null spin combination (slope
    $-2.30$ against the exact $-2.29$) with a $15\,\%$ chirp-mass admixture
    in the scaled coordinates; exponent
    $3.988\pm0.004$ over the perturbative window ($\delta\le132$),
    $3.949\pm0.015$ over all clean points; beyond $\delta\simeq50$ the ratio
    falls below unity as the local expansion ceases to describe the fit.
    }
    \label{fig:scaling_prefactor_validity}
\end{figure}

\begin{figure}[tbp]
    \centering
    \includegraphics[width=\textwidth]{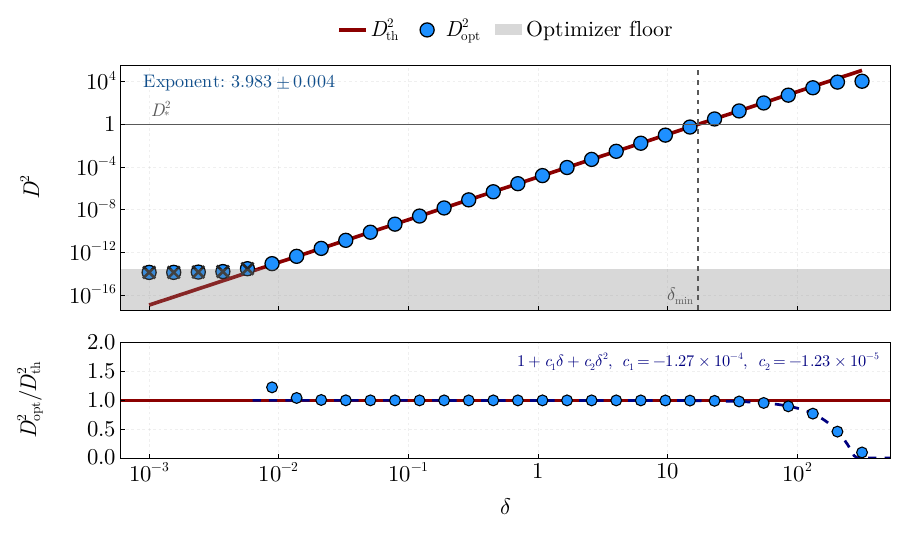}
    \caption{
    Light seed B5 along the mixed chirp-mass and spin displacement
    direction $u\propto(0,0.6,0,0,-0.55,-0.55)$, computed with the
    observation window of (\ref{eq:observation_window}), in the layout of
    Figure~\ref{fig:scaling_diagonal_masstime}: exponent $3.983\pm0.004$
    over the perturbative window ($\delta\le132$), $3.923\pm0.028$ over all
    clean points, $\delta_{\min}=17$; beyond $\delta\simeq50$ the ratio
    falls below unity as at B2$'$.
    }
    \label{fig:scaling_seed_window}
\end{figure}

The sweeps test the central prediction: after optimizing over the single
source, the squared distance scales as $\delta^4$ with the prefactor
$K_{\mathrm{num}}(u)/16$. Seven directions at six base points are reported
(Figures~\ref{fig:scaling_diagonal_masstime}--\ref{fig:scaling_seed_window});
directions are written in the scaled
coordinates $(D_L,\mathcal{M},t_c,\phi_c,\chi_1,\chi_2)$ of
Section~\ref{subsec:numerical-waveform-detector}, always with a vanishing
amplitude component, so that each sweep is a shape deformation.
Each panel shows the optimized distance against the parameter-free
prediction over thirty separations from $10^{-3}$ to $10^{2.5}$ Fisher
$\sigma$, the automatically detected optimizer floor, the threshold
$D_*^2=1$ with the corresponding $\delta_{\min}$, and, in the strip beneath
its main axis, the ratio $D^2_{\mathrm{opt}}/D^2_{\mathrm{th}}$ that exposes
the prefactor and the onset of higher-order corrections; the quoted
exponent is the fit over the perturbative window. The remaining direction,
the mixed chirp-mass and spin displacement of the light seed B5, is
computed with the observation window of (\ref{eq:observation_window}): at
that base point the confusion geometry along the mass and spin directions
is set by the start of the observation
(Section~\ref{subsec:numerical-waveform-detector}), and only the windowed
sweep describes a one-year observation.

To test the tensorial character of the scaling law without privileging any coordinate axis, we first examine a simultaneous displacement in all five shape parameters at base point B1 (Figure~\ref{fig:scaling_diagonal_masstime}, upper panel), setting the amplitude component to zero. The second source is displaced simultaneously in chirp mass, coalescence time, phase, and both spins. The numerically optimized residuals follow the theoretical $\delta^4$ curve across fifteen orders of magnitude in $D^2$ above the optimizer floor, yielding a fitted exponent of $3.995\pm0.002$ with a prefactor ratio consistent with unity throughout the clean range: the ratio stays within $0.4\,\%$ of unity up to the final separation on the grid, $\delta=316$, showing that no breakdown occurs within this range ($c_1=8.8\times10^{-7}$); the distinguishability scale is $\delta_{\min}=133$.

In gravitational-wave catalogues, the correlation between chirp mass and coalescence time constitutes a primary degeneracy: a slightly heavier binary merging slightly later produces an almost identical phase evolution. We evaluate this correlated displacement for the massive binary B2 (Figure~\ref{fig:scaling_diagonal_masstime}, lower panel). The fitted exponent is $3.994\pm0.002$ ($c_1=-1.0\times10^{-6}$); the ratio stays within $1\,\%$ of unity up to $\delta=132$ and within $3.2\,\%$ up to $\delta=316$, with $\delta_{\min}=55$. Because B2 is the base point for several confusion maps below, this sweep verifies the curvature inverted in those maps.

For the strongly anti-aligned binary B3, we examine coupled displacements in phase and both spins (Figure~\ref{fig:scaling_spin_time}, upper panel). Although this displacement enters through the 1.5PN spin--orbit coefficient rather than the Newtonian chirp, the quartic law holds over eighteen orders of magnitude in $D^2$, with an exponent of $3.994\pm0.001$ and $\delta_{\min}=49$; the ratio stays within $2\,\%$ of unity up to $\delta=132$ and departs by $8\,\%$ at $\delta=316$, marking the onset of higher-order corrections.

A pure coalescence-time shift for the band-filling binary B4 (Figure~\ref{fig:scaling_spin_time}, lower panel) isolates the geometric effect of time translation. The sum of two identical chirps displaced by $\Delta t_c$ introduces a modulation factor $2\cos(\pi f\Delta t_c)$ across the band that cannot be absorbed by a single template once the midpoint time and amplitude are fitted. The constellation motion provides an additional contribution: shifting $t_c$ moves the emission time of each frequency along the orbit, altering the Doppler phase and antenna pattern. Isolating this orbital effect would require comparing against a frozen-detector response, which is left for future work. The fitted exponent is $3.992\pm0.003$ ($c_1=5.3\times10^{-7}$); the ratio stays within $1\,\%$ of unity up to $\delta=204$ and within $2.1\,\%$ across the entire grid.

The unequal-amplitude formulation (Section~\ref{subsec:unequal}) is tested at B2 with an amplitude ratio $q=A_2/A_1=1/2$ (twice the luminosity distance; Figure~\ref{fig:scaling_prefactor_validity}, upper panel). Initializing the optimizer at the amplitude-weighted centroid with effective amplitude $(1+q)A$, the theoretical prediction includes the harmonic-mean factor $(2q/(1+q))^2=4/9$. The ratio strip confirms this prefactor: an uncorrected equal-amplitude model would predict $9/4$, whereas the measured ratio $D^2(q{=}1/2)/D^2(q{=}1)$ reproduces $4/9$ to within $10^{-4}$ across all separations where both sweeps follow the leading-order law within $10^{-3}$. The fitted exponent is $3.992\pm0.003$.

We examine the limits of the perturbative expansion at base point B2$'$ along a near-null direction: a spin displacement $(\Delta\chi_1,\Delta\chi_2)\propto(0.4,-0.92)$ (slope $-2.30$, within $0.5\%$ of the exact 1.5PN null value $-2.29$) combined with a chirp-mass offset of $0.15$ in scaled coordinates (Figure~\ref{fig:scaling_prefactor_validity}, lower panel). The offset from the null line and the mass admixture keep the curvature finite while confining the secondary source to physical spin values. Because this direction is soft---$g(u,u)$ is six to eight orders of magnitude smaller than for generic directions---one Fisher $\sigma$ corresponds to a spin displacement of $2.5\times10^{-4}$, and $K_{\mathrm{num}}$ is correspondingly elevated in Fisher units ($1.2\times10^{-4}$ compared to $10^{-8}$--$10^{-6}$ elsewhere). Over the perturbative window (24 separations up to $\delta=132$, seven times $\delta_{\min}=19$, selected by the $30\,\%$ criterion), the fitted exponent is $3.988\pm0.004$ with $c_1=-3.6\times10^{-4}$; the ratio stays within $2.5\,\%$ of unity up to $\delta=36$ and within $5.5\,\%$ up to $\delta=55$ (three times $\delta_{\min}$), falling to $0.76$ at the end of the window. Beyond $\delta\simeq50$, $D^2_{\mathrm{opt}}$ grows more gradually than $\delta^4$ and the ratio falls toward zero ($0.22$ at $\delta=316$), giving an all-points exponent of $3.949\pm0.015$. This departure does not contradict the asymptotic scaling as $\delta\to0$; rather, it indicates that at finite separation the template absorbs additional deformation beyond the leading normal projection, marking where the local expansion ceases to describe the fit. At the chosen tolerance, distinguishability is achieved earlier, at $\delta_{\min}$, well inside the window. At larger separations, $D^2_{\mathrm{opt}}$ asymptotically approaches the total power of the second source ($\rho^2\simeq4.8\times10^{5}$ at B2$'$). Along soft directions, this transition occurs at modest physical parameter separations and is traced directly by the ratio strip.

For the light binary B5, we consider a mixed chirp-mass and spin displacement where chirp mass increases while both spins decrease (Figure~\ref{fig:scaling_seed_window}). Along this direction, the chirp-mass and spin--orbit phase contributions share the same sign across the band and reinforce each other; the first-order variation is absorbed by the effective template as along every other direction. This case highlights the effect of the finite observation window: because a light binary spends decades in band before coalescence, an unwindowed calculation would accumulate power from pre-observation inspiral cycles that the instrument never recorded (Section~\ref{subsec:numerical-waveform-detector}). With the observation window included, the scaling law holds across seventeen orders of magnitude in $D^2$, with an exponent of $3.983\pm0.004$ across the perturbative window, $c_1=-1.3\times10^{-4}$, and $\delta_{\min}=17$. The normal curvature is the largest in our campaign ($K_{\mathrm{num}}=1.9\times10^{-4}$ in Fisher units), producing the smallest $\delta_{\min}$ among all seven sweeps. The ratio stays within $2\,\%$ of unity up to $\delta=36$ (twice $\delta_{\min}$) and departs by $10\,\%$ at $\delta=85$; beyond $\delta\simeq50$, it displays the same departure from the quartic law as B2$'$, reaching $0.77$ at the end of the perturbative window ($\delta=132$).

\subsection{Confusion-zone cartography}
\label{subsec:confusion-zone-maps}

The maps use the curvature diagnostic to draw the local zone of confusion in
selected parameter planes: the set of secondary-source displacements whose
best-fit single-source residual stays below the tolerance, intersected with
the physical prior box. In a plane $(\theta^p,\theta^q)$ the directions are
sampled uniformly in the \emph{Fisher-normalized} plane, each axis measured
in its own $\sigma$ at the base point, and mapped back to parameter units:
the two axes of a plane differ by up to $10^{4}$ in $\sigma$ (the
coalescence time against the phase), and a grid uniform in parameter angle
would resolve such a zone only along its axes. Because $K_{\mathrm{num}}$
is even in $u$, the curvature is evaluated on 4096 base directions of the
half-circle and mirrored, so the central symmetry of the mathematical
boundary is exact by construction; the sampling is refined adaptively
wherever the boundary radius varies by more than a fixed ratio between
neighbouring directions. Each direction gives the mathematical radius
$r_{\mathrm{math}}=(16D_*^2/K_{\mathrm{num}})^{1/4}$, capped at the distance
$r_{\mathrm{box}}$ to the prior box along that direction (the phase
deviation capped at $|\Delta\phi_c|\le\pi$; the waveform is periodic in
$\phi_c$ with period $4\pi$, a shift by $2\pi$ reversing the sign of the odd
harmonics of (\ref{eq:psi_k_numerical_model}), and no zone approaches
either bound, the phase half-extents staying below $0.4\,\mathrm{rad}$);
wherever consecutive
directions straddle a capping transition the exact crossover direction
$r_{\mathrm{math}}=r_{\mathrm{box}}$ is located by bisection and inserted
as a boundary vertex, so that zone corners at prior walls are exact rather
than chamfered. Every boundary point is flagged as curvature-limited or
prior-limited, and the prior-limited fraction of sampling angle is recorded
per map. Each zone is a local chart around one reference source and depends
on its tangent basis and curvature, on the model and response, and on the
tolerance; the information it carries is the anisotropy of the boundary.

Two caveats regarding these maps warrant mention. First, each boundary is obtained by inverting the leading-order quartic law direction by direction rather than computing level sets of the full optimized distance. While the sweeps confirm quartic scaling at each base point, exact central symmetry is a property of the leading-order quartic form that the full distance may lose at finite separation. Second, the zones are conditional on the parameters held fixed in the template family (mass ratio, sky position, inclination, polarization): freeing additional parameters enlarges the single-source manifold and can only widen the confusion zones. Direct fits at predicted boundary points will evaluate the precision of this leading-order inversion, as noted in Section~\ref{sec:conclusions}.

\begin{figure}[tbp]
    \centering
    \includegraphics[width=\textwidth]{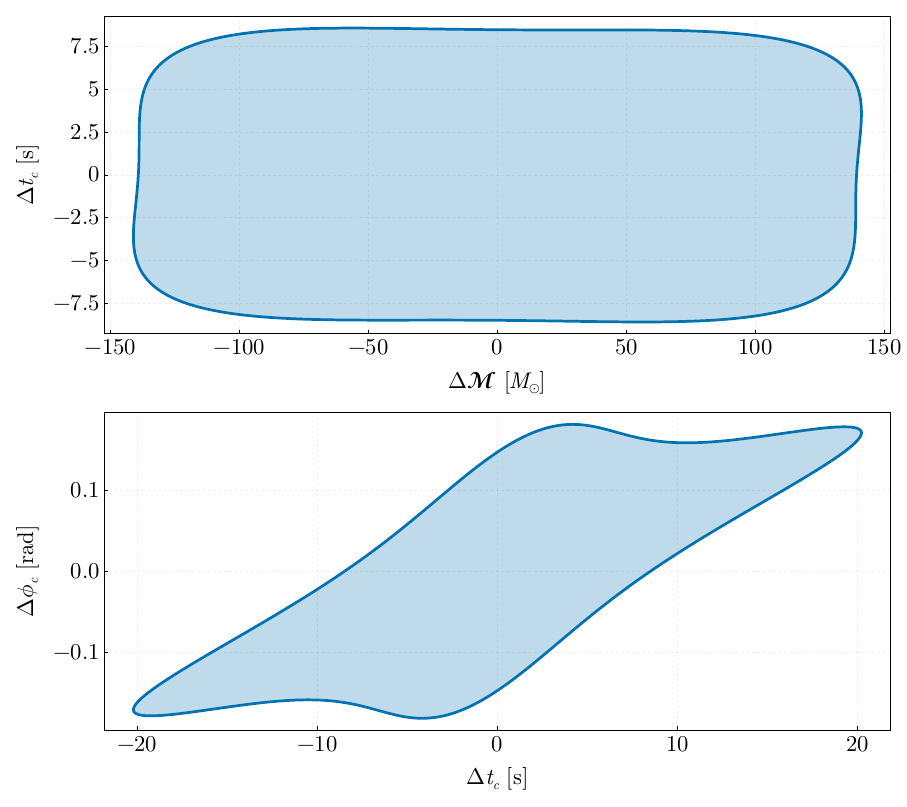}
    \caption{
    Confusion zones of the heavy base point B2 in the mass--time and
    time--phase planes, in physical units; in each panel the first-named
    parameter is on the horizontal axis. The filled region is bounded by
    the curvature-limited boundary (solid blue); both zones lie inside the
    physical priors. Upper panel: chirp mass and coalescence time,
    half-extents $140\,M_\odot$ and
    $8.6\,\mathrm{s}$; the zone is centrally symmetric by construction, while
    the odd cross term of $K(u)$ shears it, so it is not reflection-symmetric
    about the axes. Lower panel: coalescence time and phase,
    $20\,\mathrm{s}$ and $0.18\,\mathrm{rad}$; the elongation follows the
    partial compensation of a constant phase shift by a linear-in-frequency
    phase tilt.
    }
    \label{fig:zones_heavy_masstime_timephase}
\end{figure}

\begin{figure}[tbp]
    \centering
    \includegraphics[width=\textwidth]{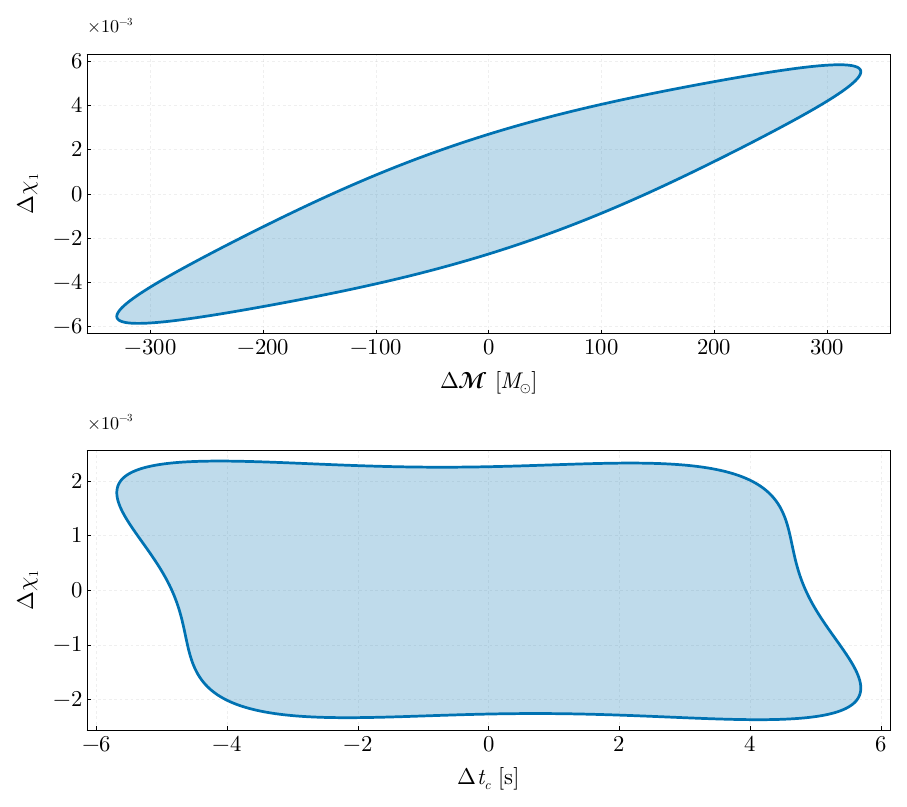}
    \caption{
    Confusion zones involving the spin $\chi_1$ at the heavy base points, in
    the layout of Figure~\ref{fig:zones_heavy_masstime_timephase}. Upper
    panel: B2, chirp mass and $\chi_1$, half-extents
    $330\,M_\odot$ and $5.9\times10^{-3}$,
    the curved mass--spin compensation of the 1.5PN phasing. Lower panel:
    B3, coalescence time and $\chi_1$, $5.7\,\mathrm{s}$ and
    $2.4\times10^{-3}$, the spin--orbit phase twist trading against arrival
    time at strong anti-alignment.
    }
    \label{fig:zones_heavy_spin}
\end{figure}

Figures~\ref{fig:zones_heavy_masstime_timephase}
and~\ref{fig:zones_heavy_spin} collect the four planes at the heavy base
points B2 and B3. All four zones are curvature-limited over every sampled
direction: at these signal-to-noise ratios the distinguishability scales are
tens of Fisher $\sigma$, i.e.\ $10^{-6}$--$10^{-2}$ in the scaled
parameters, one to five orders of magnitude within the physical boundaries. The mass--time
zone (Figure~\ref{fig:zones_heavy_masstime_timephase}, upper panel) is
compact in the time direction (a coalescence-time
displacement of some ten seconds is resolved, against a chirp-mass
displacement of a hundred and forty solar masses), while the time--phase zone
(lower panel) is elongated along the direction in which a constant phase shift
compensates a frequency-linear tilt over the band, the compensation being
imperfect because of the subdominant harmonics and the orbital Doppler
modulation. The mass--spin zone (Figure~\ref{fig:zones_heavy_spin}, upper panel) is
the widest in relative terms:
the spin--orbit term partially compensates a chirp-mass change over a
finite frequency interval. The time--spin zone at B3 (lower panel) shows the
same trade-off carried by the arrival time at strong anti-alignment.

\begin{figure}[tbp]
    \centering
    \includegraphics[width=\textwidth]{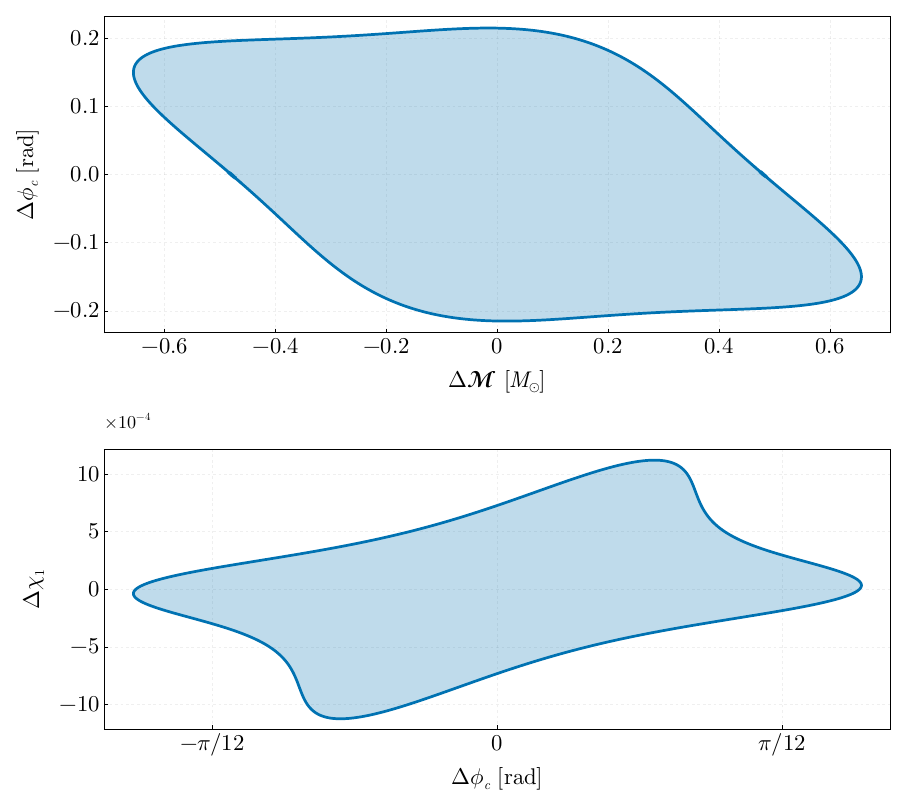}
    \caption{
    Confusion zones at the two light base points, computed with the
    observation window of (\ref{eq:observation_window}), in the layout of
    Figure~\ref{fig:zones_heavy_masstime_timephase}.
    Upper panel: B4, chirp mass and phase, half-extents
    $0.66\,M_\odot$ and $0.21\,\mathrm{rad}$.
    Lower panel: B5, phase and $\chi_1$, $0.33\,\mathrm{rad}$ and
    $1.1\times10^{-3}$. Without the window the chirp-mass half-extent of
    the upper panel is $2.8$ times and the spin half-extent of the lower
    panel $1.6$ times smaller,
    the difference being the decades of pre-observation inspiral that the
    unwindowed band would attribute to the measurement; the phase extents
    do not change.
    }
    \label{fig:zones_low_mass}
\end{figure}

At the two light base points (Figure~\ref{fig:zones_low_mass}) the signal
fills the band, and the finite observation matters: the chirp-mass and
spin information that the unwindowed band would collect from inspiral
emitted years before the observation is removed by the window, which
widens the zone along those axes by factors of $2.8$ and $1.6$ and leaves the
phase axes unchanged. Both zones remain curvature-limited; their outlines,
a bulge along the direction in which a phase shift compensates a
chirp-mass or spin change, are the low-mass counterparts of the time--phase zone of
Figure~\ref{fig:zones_heavy_masstime_timephase} (lower panel) and of the
mass--spin zone of Figure~\ref{fig:zones_heavy_spin} (upper panel).

\begin{figure}[tbp]
    \centering
    \includegraphics[width=0.86\textwidth]{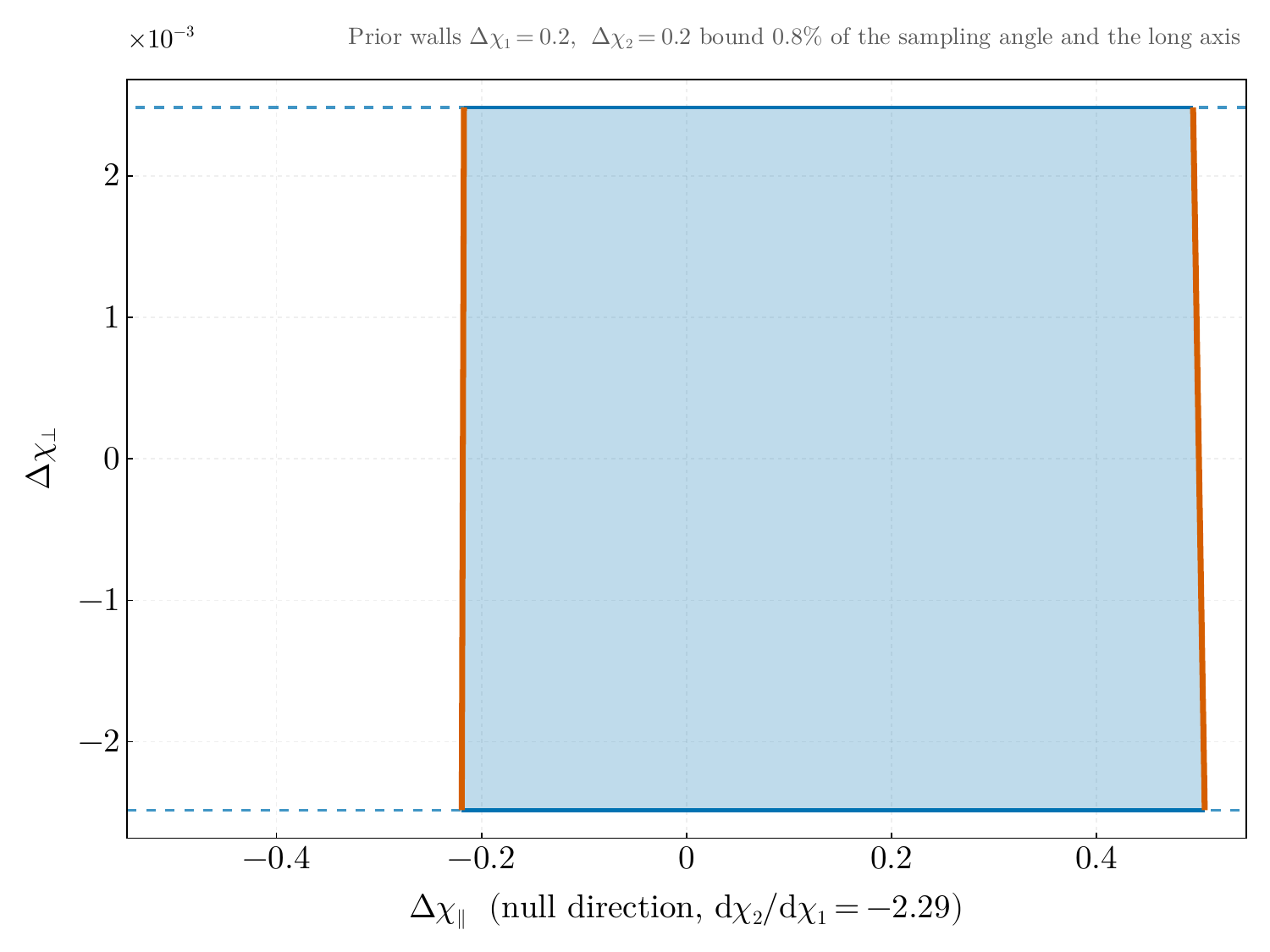}
    \caption{
    Confusion zone of B2 in the spin plane, drawn in the frame of the exact
    1.5PN null direction ($\mathrm{d}\chi_2/\mathrm{d}\chi_1=-2.29$ at
    $\eta=0.2222$): abscissa along the null line, ordinate transverse to it, on
    independent scales. Transverse to the null line the zone is
    curvature-limited with half-width $2.5\times10^{-3}$; along it the
    curvature vanishes and the zone runs to the spin walls $\chi_1=1$ and
    $\chi_2=1$ (orange segments), which bound its whole long axis although they
    subtend under $1\,\%$ of the sampling angle.
    }
    \label{fig:zone_spin_needle}
\end{figure}

The spin--spin plane (Figure~\ref{fig:zone_spin_needle}) is the direct image
of the rank deficiency of Section~\ref{subsec:numerical-waveform-detector}. Along the null combination the
waveform does not change at all, since the spins enter only through
$\beta$: the confusion is exact to every order and not only at the quartic
one, $K_{\mathrm{num}}$ vanishes and the mathematical radius diverges;
transverse to it the zone is curvature-limited with a half-width of $2.5\times10^{-3}$ in spin. The operational zone is therefore
a needle: a band of that width along the null line, terminated only by the
physical spin priors. Drawn in parameter units the band would be an edge-on
line, so the figure rotates it into the frame of the null direction. The
prior-limited fraction of sampling angle is below $1\,\%$, since few
directions hit the walls, yet the walls set the whole long axis of the zone.
This is the only prior-limited boundary of the campaign and it coincides
with the exact degeneracy of the model. The null direction is not prescribed explicitly;
it emerges directly from the degenerate tangent basis. Physical corrections that break the
degeneracy (such as 2PN spin--spin coupling or orbital precession) would close
the needle into a compact, curvature-limited zone.

Confusion is thus strongly anisotropic, exhibiting two distinct regimes:
curvature-limited boundaries shaped by the extrinsic bending of the waveform
manifold, and prior-limited segments along exact or near-exact degeneracies
where only physical bounds terminate the zone. Sweeps and maps thus serve
complementary purposes: the sweeps confirm the quartic scaling along individual
directions, while the maps invert the curvature to chart the global anisotropy
of source confusion.

\subsection{Frequency-domain anatomy of the residual}
\label{subsec:residual-spectra}

The maps summarize distinguishability geometrically but do not show how the
mismatch is distributed across the band. For a two-source signal
$\tilde\omega(f)=\tilde h(f;\theta_1)+\tilde h(f;\theta_2)$ and its best-fit single source
$\tilde h_{\mathrm{eff}}(f)=\tilde h(f;\theta_{\mathrm{eff}})$, the optimized residual is,
channel by channel, $\tilde r_{\mathrm{opt},I}(f)=\tilde\omega_I(f)-\tilde h_I(f;\theta_{\mathrm{eff}})$,
and both signal and residual are displayed as spectral densities of the
squared noise-weighted norm,
\begin{equation}
\frac{\dd\rho^2}{\dd f}\bigg|_I=\frac{4|\tilde\omega_I(f)|^2}{S_I(f)},
\qquad
\frac{\dd D^2}{\dd f}\bigg|_I=\frac{4|\tilde r_{\mathrm{opt},I}(f)|^2}{S_I(f)},
\label{eq:residual_density_definitions}
\end{equation}
so that the band integral of the residual density, summed over the
channels, is exactly the distance tested by the scaling law. The
million-bin spectra are decimated into windows uniform in $\log f$ (mean
line with min--max envelope per window; the means are lightly smoothed for
display, the envelopes are not) while every integral is evaluated on the
full grid.

\begin{figure}[tbp]
    \centering
    \includegraphics[width=\textwidth]{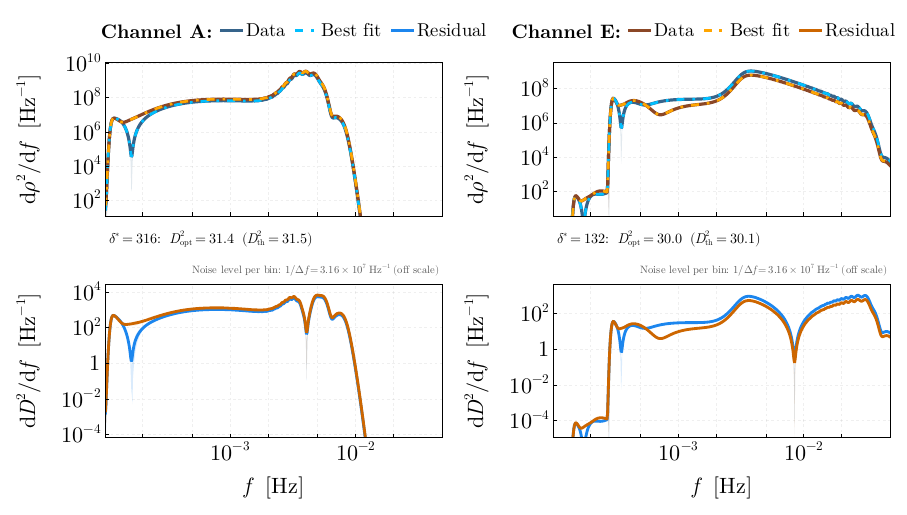}
    \caption{
    Frequency-domain comparison of the injected two-source signal and the
    optimized single-source fit, at the separation $\delta^*$ nearest
    $2\delta_{\min}$ inside the validity window (annotated). Upper panels:
    signal density $\dd\rho^2/\dd f$ of the two-source superposition and of
    the best-fit single source in the $A$ and $E$ channels; lower panels:
    the unabsorbed residual density $\dd D^2/\dd f$, whose band integral
    equals $D^2_{\mathrm{opt}}$ (annotated with the curvature prediction).
    The noise level per bin and real quadrature,
    $1/\Delta f=3.2\times10^{7}\,\mathrm{Hz}^{-1}$ in the density units of
    the panels (twice that per complex bin and channel), lies three to six
    orders of magnitude above the residual densities. Left column: B1,
    five-parameter diagonal direction, $\delta^*=316$,
    $D^2_{\mathrm{opt}}=31.44$ against $31.53$ predicted. Right column: B4,
    coalescence-time shift, $\delta^*=132$, $30.03$ against $30.15$; the
    beating of the two chirps at period $1/\Delta t_c$ tightens towards high
    frequency on the logarithmic axis. The steep rise of the B4 spectra at
    $0.3\,\mathrm{mHz}$ is the frequency the source emits when the
    observation starts: below it the window of (\ref{eq:observation_window})
    removes the signal, above about twice that frequency it is
    inconsequential.
    }
    \label{fig:residual_spectra}
\end{figure}

Figure~\ref{fig:residual_spectra} illustrates this mechanism directly in the frequency domain. Across both detector channels, the best-fit single template matches the two-source superposition over the entire band, reflecting the tangent-space cancellation established in Section~\ref{sec:local}: adjusting the effective parameters eliminates the tangential component of the two-source perturbation. The remaining irreducible residual corresponds to the normal component, $r_{\mathrm{opt}}\simeq\tfrac{A}{4}\,b_{ij}\delta\theta^i\delta\theta^j$ at leading order, whose squared norm yields the quartic scaling law. At separation $\delta^*$, the band integral of the residual density reproduces the independently minimized distance $D^2_{\mathrm{opt}}$ by construction, agreeing with the curvature prediction to within $0.4\,\%$ in both cases (consistent with the fitted $c_1\delta$ correction). Rather than point-to-point numerical scatter, the residual forms a coherent spectral pattern across the band, including interference nodes between nearby chirps in the right column. Pointwise, the residual density lies three to six orders of magnitude below the detector noise floor across all $10^{6}$ frequency bins. Distinguishability is governed not by localized spectral peaks, but by the coherent accumulation of residual power across the full bandwidth relative to the threshold $D_*$, with signal loudness entering solely as an overall scale factor.

\section{Conclusions}\label{sec:conclusions}

We have formulated the problem of distinguishing overlapping gravitational-wave signals from single-source templates in terms of the extrinsic geometry of the waveform manifold embedded in the detector Hilbert space $\mathcal{H}$.
Although the intrinsic Riemannian geometry defined by the Fisher metric $g_{ij}(\theta)=\braket{\partial_i h(\theta)|\partial_j h(\theta)}$ governs local parameter uncertainties in Gaussian noise, it cannot determine whether an unresolved composite signal is absorbed by an effective single template.

When two sources are closely spaced in parameter space, optimizing over an effective single-source template cancels all perturbations tangent to the manifold $T_{h(\theta_0)}\mathcal{M}_1$.
The leading irreducible residual lies strictly within the orthogonal normal subspace $N_{h(\theta_0)}\mathcal{M}_1$ and is governed by the second fundamental form $\mathrm{II}$ of the embedding.
This mechanism yields the quartic scaling law
\begin{equation}
D^2(\delta,u)=\frac{1}{16}\,\norm{\mathrm{II}(u,u)}^2\,\delta^4+\mathcal{O}(\delta^5),
\end{equation}
where $\norm{\mathrm{II}(u,u)}^2=\bar A^2 K(u)$ is the squared norm of the normal curvature vector along parameter direction $u$, evaluated with effective amplitude $\bar A=A$ for equal-amplitude pairs and the harmonic mean $\bar A=A_{\mathrm{harm}}=2A_1A_2/(A_1+A_2)$ for unequal amplitudes.
Prescribing an operational resolvability threshold $D_*$ establishes the directional distinguishability scale
\begin{equation}
\delta_{\min}(u)\simeq \left(\frac{16D_*^2}{\bar A^2 K(u)}\right)^{1/4}.
\end{equation}
Signal resolvability is thus intrinsically anisotropic and dictated by extrinsic curvature: directions with small $K(u)$ define locally flat submanifolds where overlapping sources merge into an effective single template, whereas directions of large extrinsic curvature generate normal residuals that allow discrimination at small parameter separations.

The numerical experiments in Section~\ref{sec:numerical-validation} confirm this geometric mechanism.
Using a six-parameter aligned-spin inspiral model with the full orbital LISA response and analytic instrumental-plus-confusion noise, we computed the noise-weighted distance-to-manifold statistic with best-fit parameters constrained to the physically admissible domain.
Across seven separation directions at six reference sources (a diagonal displacement in all five shape parameters, the chirp-mass/coalescence-time degeneracy of a massive binary, a spin--orbit displacement at strong anti-alignment, a pure coalescence-time shift, an unequal-amplitude pair, a near-null spin direction of the 1.5PN model, and a mixed chirp-mass and spin displacement of a light seed binary observed through the finite window), the measured distance follows the quartic scaling across eighteen orders of magnitude in $D^2$, with exponents between $3.983\pm0.004$ and $3.995\pm0.002$ across the perturbative window and prefactor ratios consistent with unity.
The harmonic-mean amplitude scaling is verified to better than $10^{-4}$ ($4/9$ for an amplitude ratio of one half).
Along the soft direction, the ratio of measured to predicted distance remains within $6\,\%$ of unity up to three times $\delta_{\min}$ and within $30\,\%$ up to seven times $\delta_{\min}$, charting the departure from the local expansion where higher-order corrections enter.
Seeded multi-start optimization identified no secondary minima above the optimizer floor.
Across hardware platforms, the benchmark campaign yields bit-identical results on five CPU architectures, agrees to $5\times10^{-6}$ on two others, and reproduces optimized distances to $3\times10^{-4}$ or better across GPU backends from three vendors.

Inverting the curvature scaling direction by direction and intersecting the resulting boundaries with physical priors produces operational confusion-zone maps in arbitrary parameter planes.
These boundaries represent leading-order geometric predictions for fixed extrinsic parameters; marginalizing over additional degrees of freedom can only broaden them.
Confusion is strongly anisotropic, with boundaries exhibiting two regimes: curvature-limited contours across generic directions (at tens of Fisher $\sigma$, i.e., one to five orders of magnitude within the prior boundaries) and prior-limited segments along exact degeneracies.
In the spin plane, the confusion zone forms an extended needle along the exact 1.5PN null spin trajectory, bounded solely by the spin priors.
The curvature diagnostic recovers this exact degeneracy without fine-tuning; physical corrections that break the 1.5PN degeneracy (such as 2PN spin--spin coupling or orbital precession) would regularize the needle into a closed, curvature-limited zone.
For light sources, time-domain windowing plays a substantial role: at the two band-filling base points, truncating pre-observation inspiral cycles widens the confusion zones along the chirp-mass and spin axes by factors of $2.8$ and $1.6$, respectively.

Because evaluating $K(u)$ requires only a single Jacobian and a nested-dual evaluation per direction, this diagnostic can be incorporated directly into parameter-estimation or global-fit pipelines~\cite{Littenberg2023,StrubEtAl2024,Katz2025,Deng2025}.
For each catalogued source, $K(u)$ defines the parameter domain where a secondary source remains absorbed by a single template at a chosen tolerance.
These confusion zones provide a computationally inexpensive screening metric for source multiplicity, isolating candidates where the prior astrophysical event rate predicts potential overlap and motivating targeted two-source hypothesis tests.
The deterministic distance $D^2(\omega)$ sets the expected log-likelihood increment of adding a second source at a given displacement (Section~\ref{subsec:lr}), while the directional anisotropy of the zone reveals where secondary templates must be deployed to probe multi-source hypotheses most effectively.

These findings confirm that the extrinsic geometry of the waveform manifold governs multi-source confusion within a controlled inspiral setting.
While incorporating the dominant physical effects---frequency evolution, orbital Doppler shifts, spin--orbit coupling, subdominant harmonics, time-domain windowing, and detector orbital motion---the model studied here is restricted to aligned-spin inspirals.
Natural extensions include incorporating merger and ringdown phases, extending the formulation to precessing binaries and spin-difference terms to lift the 1.5PN degeneracy, and testing the statistic across coloured noise realizations in full Bayesian sampling workflows.
In addition, direct single-source fits along predicted zone boundaries will quantify higher-order truncation errors, while systematic injection campaigns will calibrate the geometric tolerance $D_*$ against empirical false-alarm rates.
By connecting source confusion directly to the second fundamental form of the waveform manifold, extrinsic geometry provides an analytical, coordinate-invariant foundation for multi-signal resolution in gravitational-wave astronomy.

\section*{Author contributions}
\noindent\textbf{Cezar Valentin Ionescu}: Conceptualization, Formal
analysis, Writing -- original draft, Methodology, Investigation,
Validation, Visualization.\\
\noindent\textbf{Paul-Adrian Gogîță}: Software, Validation,
Investigation, Visualization, Methodology,
Conceptualization, Writing -- review \& editing.\\
\noindent\textbf{Laurențiu Ioan Caramete}: Conceptualization, Supervision, Funding
acquisition, Project administration, Writing -- review \& editing.\\
\noindent\textbf{Ana Caramete}: Supervision, Methodology, Writing -- review \&
editing.\\
\noindent All authors have read and approved the final manuscript.

\ack{The numerical simulations reported here were performed in the computing centre of the Institute of Space Science -- INFLPR Subsidiary, Măgurele, România.}

\section*{Data availability statement}
No external data sets were analysed in support of this study. All results
are reproducible with the open-source package
\textsc{CurvatureDistinguishability.jl}~\cite{CurvatureDistinguishability2026},
which includes the model, the configuration files of the campaign (physical
parameters, grid, noise model, random seeds) and the scripts that
produce the figures and tables.

\funding{Supported by the Romanian Ministry of Research, Innovation and Digitalization under the Romanian National Core Program LAPLAS VII -- contract no. 30N/2023 and by ESA PRODEX project RoLISASpace.}

\appendix

\section{Explicit local calculations for the two-source problem}\label{app:local}

\subsection{Equal-amplitude perturbative expansion}\label{app:equal}

Write
\begin{equation}
h(\theta_0)=A\,s(\nu),\qquad h(\theta_0+\delta\theta)=A\,s(\nu+\delta\theta).
\end{equation}
The two-source waveform is
\begin{align}
\omega(\delta\theta)
&=A\,s(\nu)+A\,s(\nu+\delta\theta)\nonumber\\
&=A\,s(\nu)+A\Big(s(\nu)+\partial_i s(\nu)\,\delta\theta^i+\frac12\partial_i\partial_j s(\nu)\,\delta\theta^i\delta\theta^j+\mathcal{O}(\|\delta\theta\|^3)\Big)\nonumber\\
&=2A\,s(\nu)+A\,\partial_i s(\nu)\,\delta\theta^i+\frac{A}{2}\partial_i\partial_j s(\nu)\,\delta\theta^i\delta\theta^j+\mathcal{O}(\|\delta\theta\|^3).
\end{align}
Fit with a single effective waveform
\begin{equation}
h_{\mathrm{eff}}(\alpha,\varepsilon)=(2A+\alpha)\,s(\nu+\varepsilon),
\end{equation}
with $\alpha$ and $\varepsilon^i$ small.
Taylor expand
\begin{equation}
s(\nu+\varepsilon)=s(\nu)+\partial_i s(\nu)\,\varepsilon^i+\frac12\partial_i\partial_j s(\nu)\,\varepsilon^i\varepsilon^j+\mathcal{O}(\|\varepsilon\|^3).
\end{equation}
Define the shorthand at the base point:
\begin{equation}
s:=s(\nu),\qquad s_i:=\partial_i s(\nu),\qquad s_{ij}:=\partial_i\partial_j s(\nu).
\end{equation}
Then, up to second order,
\begin{align}
h_{\mathrm{eff}}(\alpha,\varepsilon)
&=(2A+\alpha)\Big(s+s_i\varepsilon^i+\frac12 s_{ij}\varepsilon^i\varepsilon^j+\mathcal{O}(\|\varepsilon\|^3)\Big)\nonumber\\
&=2A s+\alpha s+2A s_i\varepsilon^i+\alpha s_i\varepsilon^i+A s_{ij}\varepsilon^i\varepsilon^j+\frac{\alpha}{2}s_{ij}\varepsilon^i\varepsilon^j+\mathcal{O}(\|\varepsilon\|^3,\alpha\|\varepsilon\|^2).
\end{align}

Define the residual
\begin{equation}
r(\alpha,\varepsilon;\delta\theta):=\omega(\delta\theta)-h_{\mathrm{eff}}(\alpha,\varepsilon).
\end{equation}
Using the expansions,
\begin{align}
r(\alpha,\varepsilon;\delta\theta)
&=\Big(2A s+A s_i\delta\theta^i+\frac{A}{2}s_{ij}\delta\theta^i\delta\theta^j+\mathcal{O}(\|\delta\theta\|^3)\Big)\nonumber\\
&\quad-\Big(2A s+\alpha s+2A s_i\varepsilon^i+\alpha s_i\varepsilon^i+A s_{ij}\varepsilon^i\varepsilon^j+\frac{\alpha}{2}s_{ij}\varepsilon^i\varepsilon^j+\mathcal{O}(\|\varepsilon\|^3)\Big)\nonumber\\
&=A s_i\delta\theta^i-2A s_i\varepsilon^i+\frac{A}{2}s_{ij}\delta\theta^i\delta\theta^j-A s_{ij}\varepsilon^i\varepsilon^j-\alpha s-\alpha s_i\varepsilon^i-\frac{\alpha}{2}s_{ij}\varepsilon^i\varepsilon^j\nonumber\\
&\qquad+\mathcal{O}(\|\delta\theta\|^3,\|\varepsilon\|^3,\alpha\|\varepsilon\|^2).
\end{align}

We solve the stationarity conditions order by order by introducing a bookkeeping parameter $\lambda$:
\begin{equation}
\delta\theta^i=d^i\lambda,\qquad
\varepsilon^i(\lambda)=\varepsilon^{(1)i}\lambda+\varepsilon^{(2)i}\lambda^2+\cdots,\qquad
\alpha(\lambda)=\alpha^{(1)}\lambda+\alpha^{(2)}\lambda^2+\cdots.
\end{equation}
Collect orders in $\lambda$:

\paragraph{Order $\lambda$.}
At order $\lambda$,
\begin{equation}
r^{(1)}(\lambda)=\lambda\Big(A s_i(d^i-2\varepsilon^{(1)i})-\alpha^{(1)}s\Big).
\end{equation}
At this order the residual lies in the tangent span $\mathrm{span}\{s,s_i\}$, and orthogonality to the tangent basis forces the coefficients to vanish:
\begin{equation}
d^i-2\varepsilon^{(1)i}=0\quad \forall i,
\qquad
\alpha^{(1)}=0.
\end{equation}
Thus
\begin{equation}
\varepsilon^{(1)i}=\frac12 d^i=\frac12\,\delta\theta^i,\qquad \alpha^{(1)}=0,
\end{equation}
and in particular $\varepsilon=\mathcal{O}(\delta\theta)$ while $\alpha=\mathcal{O}(\delta\theta^2)$.

\paragraph{Order $\lambda^2$.}
Expanding $r(\alpha,\varepsilon;\delta\theta)$ to second order and substituting the first-order solution $\varepsilon^{(1)i}=\frac12 d^i$, $\alpha^{(1)}=0$, the quadratic template term contributes $-A s_{ij}\varepsilon^{(1)i}\varepsilon^{(1)j}\lambda^2 = -\frac{A}{4}s_{ij}d^i d^j\lambda^2$, while cross terms involving $\alpha$ vanish at this order since $\alpha=\mathcal{O}(\lambda^2)$.
Combining this with the source term $\frac{A}{2}s_{ij}d^i d^j\lambda^2$ gives
\begin{equation}
r^{(2)}=\frac{A}{4}s_{ij}d^i d^j-2A s_i\varepsilon^{(2)i}-\alpha^{(2)}s.
\end{equation}
From here on we set $\lambda=1$, so that $d^i=\delta\theta^i$ and $r^{(n)}$, $\varepsilon^{(n)}$ and $\alpha^{(n)}$ denote the terms homogeneous of degree $n$ in $\delta\theta$.

Now decompose the second derivative into tangent and normal parts, with $b_{ij}$ of (\ref{eq:b_def}):
\begin{equation}
s_{ij}=P_T[s_{ij}]+P_N[s_{ij}]
=P_T[s_{ij}]+b_{ij}.
\end{equation}
Then
\begin{align}
r^{(2)}
&=\frac{A}{4}b_{ij}\delta\theta^i\delta\theta^j
+\Big(\frac{A}{4}P_T[s_{ij}]\delta\theta^i\delta\theta^j-2A s_i\varepsilon^{(2)i}-\alpha^{(2)}s\Big).
\end{align}
By construction the first term is normal; the bracketed term is tangent and can be cancelled by suitable (and under non-degeneracy, unique) choices of $\varepsilon^{(2)}$ and $\alpha^{(2)}$.
Therefore the irreducible normal residual begins at second order:
\begin{equation}
r_N(\delta\theta)=r_N^{(2)}(\delta\theta)+\mathcal{O}(\|\delta\theta\|^3),
\qquad
r_N^{(2)}(\delta\theta)=\frac{A}{4}\,b_{ij}\delta\theta^i\delta\theta^j.
\end{equation}
Consequently,
\begin{align}
D^2(\omega(\delta\theta))
&=\norm{r_N(\delta\theta)}^2
=\norm{r_N^{(2)}(\delta\theta)}^2+\mathcal{O}(\|\delta\theta\|^5)\nonumber\\
&=\frac{A^2}{16}\,\ip{b_{ij}}{b_{k\ell}}\,
\delta\theta^i\delta\theta^j\delta\theta^k\delta\theta^\ell+\mathcal{O}(\|\delta\theta\|^5).
\end{align}
With $C_{ijk\ell}$ of (\ref{eq:quartic_tensor}),
\begin{equation}
D^2(\omega(\delta\theta))=\frac{A^2}{16}\,C_{ijk\ell}\,\delta\theta^i\delta\theta^j\delta\theta^k\delta\theta^\ell+\mathcal{O}(\|\delta\theta\|^5).
\end{equation}
For a unit direction $u$ with respect to the shape-family metric $G_{ij}:=\ip{s_i}{s_j}=g_{ij}/A^2$ and $\delta\theta^i=\delta u^i$ (this split of $\delta\theta$ into $\delta$ and $u$ differs from the $g$-normalized one of Section~\ref{subsec:equal} by the factor $A$, which cancels between $K(u)$ and $\delta^4$),
\begin{equation}
D^2(\delta,u)=\frac{A^2}{16}\,K(u)\,\delta^4+\mathcal{O}(\delta^5),
\qquad
K(u):=C_{ijk\ell}u^iu^ju^ku^\ell.
\end{equation}
Finally, by (\ref{eq:II_vs_b}) $A\,b_{ij}$ is the second fundamental form of $\mathcal{M}_1$ at the source point $h(\theta_0)$, so $A^2K(u)=\norm{\mathrm{II}_{h(\theta_0)}(u,u)}^2$ as used in the main text.

\subsection{Unequal-amplitude perturbative expansion}\label{app:unequal}

Adopt the same decomposition $h(\theta)=A\,s(\nu)$ but consider two sources with unequal amplitudes:
\begin{equation}
\theta_1=(A_1,\nu),
\qquad
\theta_2=(A_2,\nu+\delta\theta),
\qquad
A_\Sigma:=A_1+A_2.
\end{equation}
The two-source waveform is
\begin{align}
\omega(\delta\theta)
&=A_1 s(\nu)+A_2 s(\nu+\delta\theta)\nonumber\\
&=(A_1+A_2)s + A_2 s_i\delta\theta^i+\frac{A_2}{2}s_{ij}\delta\theta^i\delta\theta^j+\mathcal{O}(\|\delta\theta\|^3).
\end{align}
Fit it with a single effective source
\begin{equation}
h_{\mathrm{eff}}(\alpha,\varepsilon)=(A_\Sigma+\alpha)\,s(\nu+\varepsilon),
\end{equation}
and expand to second order:
\begin{equation}
h_{\mathrm{eff}}=A_\Sigma s+\alpha s+A_\Sigma s_i\varepsilon^i+\alpha s_i\varepsilon^i+\frac{A_\Sigma}{2}s_{ij}\varepsilon^i\varepsilon^j+\mathcal{O}(\|\varepsilon\|^3,\alpha\|\varepsilon\|^2).
\end{equation}
Define the residual:
\begin{align}
r(\alpha,\varepsilon;\delta\theta)
&=\omega(\delta\theta)-h_{\mathrm{eff}}(\alpha,\varepsilon)\nonumber\\
&=A_2 s_i\delta\theta^i-A_\Sigma s_i\varepsilon^i-\alpha s
+\frac{A_2}{2}s_{ij}\delta\theta^i\delta\theta^j-\frac{A_\Sigma}{2}s_{ij}\varepsilon^i\varepsilon^j-\alpha s_i\varepsilon^i
+\mathcal{O}(\|\delta\theta\|^3,\|\varepsilon\|^3,\alpha\|\varepsilon\|^2).
\end{align}

Expand the minimizer in homogeneous pieces:
\begin{equation}
\varepsilon^i=\varepsilon^{(1)i}+\varepsilon^{(2)i}+\mathcal{O}(\|\delta\theta\|^3),
\qquad
\alpha=\alpha^{(1)}+\alpha^{(2)}+\mathcal{O}(\|\delta\theta\|^3),
\end{equation}
where $\varepsilon^{(n)}$ and $\alpha^{(n)}$ are homogeneous of degree $n$ in $\delta\theta$.
Accordingly,
\begin{equation}
r=r^{(1)}+r^{(2)}+\mathcal{O}(\|\delta\theta\|^3),
\end{equation}
with
\begin{equation}
r^{(1)}=A_2 s_i\delta\theta^i-A_\Sigma s_i\varepsilon^{(1)i}-\alpha^{(1)}s,
\end{equation}
and the second-order contribution
\begin{equation}
r^{(2)}=\frac{A_2}{2}s_{ij}\delta\theta^i\delta\theta^j-\frac{A_\Sigma}{2}s_{ij}\varepsilon^{(1)i}\varepsilon^{(1)j}-A_\Sigma s_i\varepsilon^{(2)i}-\alpha^{(2)}s-\alpha^{(1)}s_i\varepsilon^{(1)i}.
\end{equation}
At the minimizer the residual is orthogonal to the tangent space at the fitted point; since the first-order residual vanishes, expanding this condition about the base point $(A_\Sigma,\nu)$ gives orthogonality of $r^{(1)}$ and $r^{(2)}$ to the tangent basis at the base point.
The tangent vectors are
\begin{equation}
t_0=\frac{\partial h}{\partial A}\Big|_{(A_\Sigma,\nu)}=s,
\qquad
t_i=\frac{\partial h}{\partial \nu^i}\Big|_{(A_\Sigma,\nu)}=A_\Sigma s_i.
\end{equation}
(If convenient, one may normalize $s$ so that $\ip{s}{s}=1$ at the base point; otherwise the same conclusions follow with a rescaled amplitude coordinate.)

\paragraph{First order.}
The first-order residual lies in the base-point tangent span and is
orthogonal to that span, so it vanishes. On the identifiable subspace,
linear independence of the tangent vectors then gives $\alpha^{(1)}=0$.
The shape stationarity conditions reduce to
\begin{equation}
0=\ip{r^{(1)}}{A_\Sigma s_j}
=A_\Sigma\left(A_2\,\ip{s_i}{s_j}\delta\theta^i-A_\Sigma\,\ip{s_i}{s_j}\varepsilon^{(1)i}\right),
\end{equation}
so, assuming the shape-family metric $G_{ij}=\ip{s_i}{s_j}$ is invertible on the identifiable subspace,
\begin{equation}
\varepsilon^{(1)i}=\frac{A_2}{A_\Sigma}\,\delta\theta^i.
\end{equation}

\paragraph{Second order.}
Substitute $\alpha^{(1)}=0$ and $\varepsilon^{(1)}$ into $r^{(2)}$:
\begin{align}
r^{(2)}
&=\frac12\left(A_2-\frac{A_2^2}{A_\Sigma}\right)s_{ij}\delta\theta^i\delta\theta^j
-A_\Sigma s_i\varepsilon^{(2)i}-\alpha^{(2)}s\nonumber\\
&=\frac{A_1A_2}{2A_\Sigma}\,s_{ij}\delta\theta^i\delta\theta^j
-A_\Sigma s_i\varepsilon^{(2)i}-\alpha^{(2)}s,
\end{align}
using $A_2-\frac{A_2^2}{A_\Sigma}=\frac{A_1A_2}{A_\Sigma}$.
Decompose $s_{ij}$ into tangent and normal parts, with $b_{ij}$ of (\ref{eq:b_def}).
Then, as in the equal-amplitude case, the tangent terms can be cancelled by suitable $\varepsilon^{(2)}$ and $\alpha^{(2)}$,
while the normal component cannot be cancelled. The irreducible normal residual at second order is therefore
\begin{equation}
r_N^{(2)}(\delta\theta)=\frac{A_1A_2}{2(A_1+A_2)}\,b_{ij}\delta\theta^i\delta\theta^j.
\end{equation}
Therefore,
\begin{align}
D^2(\omega(\delta\theta))
&=\norm{r_N(\delta\theta)}^2
=\norm{r_N^{(2)}(\delta\theta)}^2+\mathcal{O}(\|\delta\theta\|^5)\nonumber\\
&=\left(\frac{A_1A_2}{2(A_1+A_2)}\right)^2
C_{ijk\ell}\,\delta\theta^i\delta\theta^j\delta\theta^k\delta\theta^\ell
+\mathcal{O}(\|\delta\theta\|^5),
\end{align}
with $C_{ijk\ell}$ of (\ref{eq:quartic_tensor}).
Defining the harmonic-mean amplitude
\begin{equation}
A_{\mathrm{harm}}:=\frac{2A_1A_2}{A_1+A_2},
\end{equation}
one has $\left(\frac{A_1A_2}{2(A_1+A_2)}\right)^2=\frac{A_{\mathrm{harm}}^2}{16}$, and thus
\begin{equation}
D^2(\omega(\delta\theta))=\frac{A_{\mathrm{harm}}^2}{16}\,
C_{ijk\ell}\,\delta\theta^i\delta\theta^j\delta\theta^k\delta\theta^\ell+\mathcal{O}(\|\delta\theta\|^5).
\end{equation}
Equivalently, for $\delta\theta^i=\delta u^i$ with $G_{ij}u^iu^j=1$ and $K(u):=C_{ijk\ell}u^iu^ju^ku^\ell$,
\begin{equation}
D^2(\delta,u)=\frac{A_{\mathrm{harm}}^2}{16}\,K(u)\,\delta^4+\mathcal{O}(\delta^5).
\end{equation}

\bibliographystyle{iopart-num}
\bibliography{references}

\end{document}